\documentstyle[12pt]{article}
\newcommand{\ret}{\nonumber \\}
\newcommand{\bigno}{\bigskip\noindent}
\newcommand{\Section}[1]%
{\section{#1}\setcounter{equation}{0}%
\setcounter{theorem}{0}}
\newtheorem{theorem}{Theorem}
\newtheorem{lemma}[theorem]{Lemma}
\newtheorem{coro}[theorem]{Corollary}
\newtheorem{pro}[theorem]{Proposition}
\newtheorem{definition}[theorem]{Definition}
\newtheorem{conjecture}[theorem]{Conjecture}

\newenvironment{proof}[1]%
{\par\noindent{\em #1:\ }}%
{~\rule{2mm}{2mm}\par\bigskip}
\newcommand\calA{{\cal A}}
\newcommand\calE{{\cal E}}
\newcommand\calG{{\cal G}}
\newcommand\calH{{\cal H}}
\newcommand\calP{{\cal P}}
\newcommand\calS{{\cal S}}
\newcommand\Zd{{\bf Z}^d}
\newcommand\gen{C_\Lambda}
\newcommand\ham{H_\Lambda}
\newcommand\hampbc{H_\Lambda^{\rm p.b.c.}}
\newcommand\op{O_\Lambda}

\newcommand\opone{O_\Lambda^{(1)}}
\newcommand\optwo{O_\Lambda^{(2)}}
\newcommand\opplus{O_\Lambda^+}
\newcommand\opminus{O_\Lambda^-}
\newcommand\PL{\Phi_\Lambda}
\newcommand\Pz{\Phi_\Lambda^{(0)}}
\newcommand\PsL{\Psi_\Lambda}
\newcommand\EL{E_\Lambda}
\newcommand\TDL{{\Lambda\uparrow{\bf Z}^d}}
\newcommand\toinf{\uparrow\infty}
\newcommand\vbar{\,\,\Bigl|\,\,}
\newcommand\barOmega{\overline{\Omega}}
\newcommand{\mmax}{m_{\rm max}}
\newcommand{\bkt}[1]{(\Phi_\Lambda, #1 
\,\Phi_\Lambda)}
\newcommand{\bktz}[1]{(\Phi_\Lambda^{(0)}, #1 
\,\Phi_\Lambda^{(0)})}
\newcommand{\abs}[1]{\left|#1\right|}
\newcommand{\norm}[1]{\left\Vert#1\right\Vert}
\newcommand{\rbk}[1]{\left(#1\right)}
\newcommand{\sbk}[1]{\left[#1\right]}
\newcommand{\cbk}[1]{\left\{#1\right\}}
\newcommand{\gs}[1]{(\Phi,#1\,\Phi)}
\newcommand{\oone}{O^{(1)}}
\newcommand{\Oplus}{O^+}
\newcommand{\Ominus}{O^-}
\newcommand{\osig}[1]{O^{\sigma_{#1}}}
\newcommand{\kone}{(\oone)}
\newcommand{\kplus}{(\Oplus)}
\newcommand{\kminus}{(\Ominus)}
\newcommand{\Rx}{R_x}
\newcommand{\Qx}{Q_x}
\newcommand{\Rsx}{R_x^*}
\newcommand{\Qsx}{Q_x^*}
\newcommand{\KRx}{(\Rx)}
\newcommand{\KQx}{(\Qx)}
\newcommand{\KRsx}{(\Rsx)}
\newcommand{\KQsx}{(\Qsx)}
\newcommand{\Mk}{{M\choose k}}
\newcommand{\Ml}{{M\choose\ell}}
\newcommand{\JJ}{{2J\choose J}}
\renewcommand{\hat}{\widehat}
\renewcommand{\tilde}{\widetilde}
\newcommand{\EG}{E_{\rm G}}
\newcommand{\PLO}{\Phi_\Lambda^{(0)}}
\newcommand{\XL}{\Xi_\Lambda}
\newcommand{\AO}{A_\Omega}
\newcommand{\AOS}{(A_\Omega)^2}
\newcommand{\amp}[3]{(\Phi^{(#1)},#2\,\Phi^{(#3)})}
\newcommand{\absbar}{\Bigl|}
\newcommand{\BO}{B_\Omega}
\newcommand{\BL}{B_\Lambda}
\begin{document}
\thispagestyle{empty}
\begin{center}
{\large\bf
Symmetry Breaking and
Finite Size Effects in
Quantum Many-Body Systems\footnote{
Published in J. Stat. Phys. {\bf 76},  745--803 (1994).
Notes are added in footnote \ref{BEfn} and in the Remark of 
Section~\ref{BEsec}.
}
}
\par\bigskip
Tohru Koma\footnote{
koma@riron.gakushuin.ac.jp
}
and
Hal Tasaki\footnote{
hal.tasaki@gakushuin.ac.jp
}
\par\bigskip
{\it Department of Physics,
Gakushuin University,
Mejiro, Toshima-ku, Tokyo 171,
JAPAN}
\end{center}
\begin{abstract}
We consider a quantum many-body system
on a lattice which exhibits a 
spontaneous symmetry breaking in its infinite volume ground states, 
but in which 
the corresponding order operator does not commute with the 
Hamiltonian.
Typical examples are the Heisenberg antiferromagnet with a N\'eel
order, and the Hubbard model with a (superconducting) off-diagonal
long range order.
In the corresponding finite system, the symmetry breaking is usually 
``obscured'' by ``quantum fluctuation'' 
and one gets a symmetric ground 
state with a long range order.
In such a situation, Horsch and von der Linden proved 
that the finite system has a low-lying eigenstate
whose excitation energy is not more than of order $N^{-1}$, 
where $N$ denotes the 
number of the sites in the lattice.
Here we study the situation where the broken symmetry is a continuous
one.
For a particular set of states (which are orthogonal to the
ground state and with each other), we prove bounds for their 
energy expectation values.
The bounds establish that there exist ever increasing numbers of
low-lying eigenstates whose excitation energies are bounded by a constant
times $N^{-1}$.
A crucial feature of the particular low-lying states we 
consider is that they can be
regarded as finite-volume counterparts of the infinite volume ground
states.
By forming linear combinations of these low-lying states and the 
(finite-volume) ground state, and by taking infinite volume limits, 
we construct infinite volume ground states with explicit symmetry breaking.
We conjecture these infinite volume ground states
to be ergodic, {\em i.e.\/}, physically natural.
Our general theorems do not only shed light on the nature of
symmetry breaking in quantum many-body systems, but
provide indispensable information for numerical approaches
to these systems.
We also discuss applications of our general results to a
variety of interesting examples.
The present paper is intended to be 
accessible to the readers
without background in mathematical approaches to 
quantum many-body systems.
\par\bigskip
{\bf KEY WORDS:} Symmetry breaking; 
long range order; 
obscured symmetry breaking;
finite-size effects;
quantum fluctuation;
ground states; 
low-lying  states; 
ergodic states.
\end{abstract}
\noindent
\newpage
\tableofcontents
\newpage
\Section{Introduction}
\label{Sec1}
\subsection{Motivations}
Symmetry breaking in quantum many-body systems is a challenging 
problem in theoretical physics.
In some situations, strong ``quantum effects'' lead to phenomena 
which are hard to predict or understand from ``classical'' points of 
view.
The present paper is devoted to one of such ``quantum effects'', 
namely, how a symmetry breaking manifests itself in a {\em finite\/} 
system when the order operator and the Hamiltonian do not 
commute with each other.
The topic reflects subtlety of both quantum mechanics and 
many-body 
problems, and indeed has been a source of some confusions in the 
field\footnote{
See footnotes in Section~\ref{SecEx} 
for what the confusions are.
}.
We have tried in the present paper not only to present our new 
theorems, but 
also to review (hopefully in an accessible manner) some background 
materials 
necessary to understand the nature of the problems.

Suppose that we have a quantum many-body system which exhibits 
a spontaneous symmetry breaking in its infinite volume ground 
states. 
When the operator that measures the symmetry does not commute 
with the Hamiltonian, one encounters strong ``quantum 
fluctuation''.  
In the corresponding {\em finite\/} system, the symmetry breaking 
is usually ``obscured'' by the fluctuation, and one only gets a unique 
ground state with perfect symmetry.

An ``obscured symmetry breaking\footnote{
Even in a classical system, or a quantum system with commuting Hamiltonian
and order operator, one never observes explicit symmetry breaking in a 
finite system at a {\em finite temperature\/}.
In this sense,  ``obscured symmetry breaking'' may be regarded as 
a  common phenomenon not necessarily intrinsic to quantum systems.
Throughout the present paper, however, we use the term ``obscured symmetry
breaking'' to indicate only the (most interesting and nontrivial)
situation where a symmetry breaking in the infinite
volume becomes unobservable in a {\em ground state\/} 
of a finite system due to
quantum fluctuation.
}'' usually manifests itself in the following
two different ways.
\begin{itemize}
\item
One observes a long range order in the ground state two-point 
correlation function for the order operators.
\item
There appear eigenstates of the Hamiltonian with energies ``close'' to 
the ground state energy.
We call them ``low-lying eigenstates''.
(See Section~\ref{SecPre} for precisely how ``close'' the energies should be,
and what is the motivation for the criterion.)
\end{itemize}
Although one might be tempted to interpret these ``low-lying 
eigenstates''
as counterparts of excited states in the infinite volume system, 
some of them 
are actually ``parts'' of the infinite volume ground states.
In the infinite volume limit, some of the ``low-lying 
eigenstates'' and the unique 
ground state are linearly combined to form a set of ergodic 
ground states with explicit symmetry breaking. 
These ergodic ground states are believed to correspond to physically
realizable states in a large system at an extremely low temperature.

When a continuous symmetry is broken, there are infinitely many 
ergodic ground states in the infinite volume. 
It is then expected that the number of independent
``low-lying eigenstates'' in the 
corresponding finite system increases indefinitely as the system 
size gets larger.
The existence of such ever increasing numbers of ``low-lying states'' 
and the corresponding finite-size scaling behavior of the low-lying
spectrum of the Hamiltonian may be regarded as characteristic
features of a continuous symmetry breaking in a quantum many-body 
system.
These points
have been discussed mainly by practitioners of numerical exact 
diagonalization of quantum spin systems.
(It is not easy to list all the relevant references.
See, for example, 
\cite{KaplanHorschLinden,Kikuchi,Bernu,Azaria,Leung,Momoi1} 
and the references therein.)
But rigorous information had been lacking
except in the mean-field model 
\cite{LiebMattis,Kaiser,KaplanLinden}.
See also \cite{Anderson} for early related discussion within the 
framework of the spin wave approximation.

The purpose of the present paper is to state general 
theorems for lattice quantum many-body systems
which exhibit  ``obscured symmetry breaking''.
Our results can be roughly divided into two parts,
which are closely related with each other.

The first set of results clarifies the relation between the
above mentioned two types of manifestations of an 
``obscured symmetry breaking''.
Whenever there is a finite volume 
ground state which does not break symmetry but whose correlation 
function exhibits a certain long range order, we expect that there 
inevitably appear ``low-lying eigenstates''.
Horsch and von der Linden  \cite{HorschLinden} actually proved
that the existence of a long range order implies  
the existence of a ``low-lying eigenstate'' whose excitation energy 
is less than of order $N^{-1}$, where $N$ is the number of 
sites in the lattice.
(See Theorem~\ref{M=1theorem}.)
Our new results deal with the cases where
the long range order is related to a continuous $U(1)$ symmetry.
We construct a particular set of states which are orthogonal to
the ground state and with each other, and prove that they are indeed
``low-lying states'' in the sense of Definition~\ref{DEFlls}.
When the system has a higher $U(1)\times{\bf Z}_2$ symmetry,
we can show that there are ever increasing numbers of ``low-lying
eigenstates'' whose excitation energies are bounded from
above by a constant times $N^{-1}$.
Such finite size scaling behavior of the low-lying spectrum of 
the Hamiltonian is characteristic in a system
where an ``obscured symmetry breaking'' related to a continuous
symmetry takes place.
As far as we know, this is the first rigorous (and explicit) 
demonstration that 
ever increasing numbers of ``low-lying states'' indeed exist. 

A very important problem which we do {\em not\/}
solve in the present paper is whether such finite size scaling behavior
alone is sufficient to conclude that there is a symmetry breaking in
the infinite volume.
See \cite{Bernu,Azaria,Momoi1} for some discussions about this
problem.
See \cite{KomaTasaki} for a solution of another closely related problem of
showing the existence of a symmetry breaking when there is a long 
range order in a series of finite systems.

The second set of results of the present paper
clarifies the roles played by the
``low-lying states'' in forming infinite volume ground states
with explicit symmetry breaking.
In general we show that any translation invariant ``low-lying
states'' converge to a ground state in the infinite volume limit.
The particular  ``low-lying states'' we construct have
a crucial feature that they can be naturally regarded as
``parts'' of infinite volume ground states with explicit symmetry
breaking.
To demonstrate this fact, 
we construct (for a general class of models with a $U(1)$ symmetry)
infinite volume ground states with explicit symmetry breaking
by taking suitable linear combinations of the 
low-lying states and the (finite volume) ground state, 
and then taking infinite volume limits.
We conjecture that these ground states are ergodic, {\em i.e.\/}, are 
physically natural (infinite volume) ground states.

We also discuss applications of our general results to a variety of
concrete examples.
The examples include Heisenberg 
antiferromagnet, the Bose-Einstein condensation in the hard core Bose 
gas on a lattice, 
the superconductivity in lattice electron 
models, and the Haldane gap problem in $S=1$ quantum antiferromagnetic 
chain.
An interesting observation in the application to the Bose gas is
that, by following our general discussions, 
we are naturally led
to consider ground states which do not conserve the particle number.

We believe that these results do not only clarify the nature of 
symmetry breaking phenomena in quantum systems, 
but also provide indispensable information for numerical 
approaches to various quantum many-body systems. 

The present paper is organized as follows.
In the following Section~\ref{SecEx}, we illustrate some of the basic 
notions by studying a concrete example of the Ising model under 
a transverse magnetic field.
We have tried to make this section accessible
 to the readers who are not 
familiar with mathematical approaches to quantum many-body 
problems.
In Section~\ref{Sec2}, we state our theorems in the most general setting,
and discuss their physical consequences.
In Section~\ref{Sec3}, we discuss applications of our theorems 
to typical problems.
Sections~\ref{Sec4} and \ref{Sec5} are devoted to the proofs of our theorems. 

In three Appendices, we prove and summarize some useful 
results closely related to the main body of the paper.
In Appendix~\ref{APgs}, we discuss relations between three different 
definitions of infinite volume ground states, and show that
they are all equivalent when restricted to translation invariant 
states.
In Appendix~\ref{APergodic}, we concentrate on a system with 
spontaneously broken discrete symmetry, and present a theorem
which shows how to construct ergodic infinite volume ground
states.
In Appendix~\ref{APfluc}, we prove lemmas which characterize
fluctuations of bulk quantities.

\subsection{``Obscured symmetry breaking'' and ``low-lying states''
in a simple example}
\label{SecEx}
Before discussing general theorems, we want to make clear what 
we mean by ``obscured symmetry breaking'' and ``low-lying 
states'', and how these notions are related to phenomena of 
symmetry breaking.
For this purpose, we shall 
discuss one of the simplest models
in which one observes an ``obscured symmetry breaking'' and 
``low-lying states''.
In the course of the discussion, 
we briefly review the notions of ground states in an 
infinite system, of ergodic states, and of symmetry breaking in 
the absence of a symmetry breaking field.
Although such materials form standard background in 
mathematical approaches to quantum many-body systems, we have 
noted that they are not widely appreciated in standard physics 
literature.
Here we will try to explain basic physical ideas rather than 
developing 
precise mathematical formalism.
Mathematical details will be supplied in the following sections.

Consider the $d$-dimensional $L\times\cdots\times L$ hypercubic 
lattice $\Lambda \subset {\bf Z}^d$, and impose periodic boundary 
conditions.
We define the $S=1/2$ 
spin system on $\Lambda$ with the Hamiltonian
\begin{equation}
\ham = -\sum_{\langle x,y \rangle} S^{(3)}_{x}S^{(3)}_{y} - 
B\sum_xS^{(1)}_{x},
\label{Ising1}
\end{equation}
where the first sum is over nearest neighbor pairs of sites in 
$\Lambda$, the magnetic 
field satisfies $B\ge0$, and
${\bf S}_x=(S^{(1)}_x,S^{(2)}_x,S^{(3)}_x)$ denote the $S=1/2$ spin 
operators at site $x$.
The model is known as the Ising model under transverse magnetic 
field.

The ground state of the Hamiltonian (\ref{Ising1}) is known to 
exhibit a phase transition as the transverse field $B$ is varied.
This is most clearly seen from the following behavior of the order 
parameter $m(B)$.
Let $\Pz(B,B')$ be the normalized ground state of the Hamiltonian 
$\ham-B'\op$, where $\op$ is the order operator
\begin{equation}
\op=\sum_{x\in\Lambda}S^{(3)}_x.
\label{opIsing}
\end{equation}
The field $B'$ is usually called symmetry breaking field.
Define the order parameter by
\begin{equation}
m(B):=\lim_{B'\downarrow0}\lim_\TDL\frac{1}{N}
(\Pz(B,B'),\op\,\Pz(B,B')),
\end{equation}
where $N=L^d$ is the number of sites in $\Lambda$.
Throughout the present paper, the symbol $:=$ signifies definition.
It can be proved that, for a fixed dimension $d(=1,2,3,\ldots$), the order operator 
satisfies $m(B)=0$ for sufficiently large $B$, and $m(B)>0$ for 
sufficiently small $B$.
In the latter case, the global up-down symmetry of the system
is spontaneously 
broken in the infinite volume ground state.
(Of course we mean the positive ({\em resp.\/}, negative)
direction in the third axis by up ({\em resp.\/}, down).)

Let us see how this symmetry breaking manifests itself in {\em 
finite} systems.
When $B=0$, the model is nothing but the classical Ising model.
The Hamiltonian (\ref{Ising1}) has two ground states 
$\PL^+$ and $\PL^-$, in which all the spins are pointing up and 
down, 
respectively.
The ground states are ordered, and break the up-down
symmetry of the Hamiltonian.

When $B>0$, we encounter ``quantum fluctuation''.
By using the 
Perron-Frobenius theorem,
as in Marshall \cite{Marshall} and
in Lieb and Mattis \cite{LiebMattis},
one finds that the ground state $\Pz(B)$ 
of the 
Hamiltonian (\ref{Ising1}) is unique for an arbitrary finite $L$.
Hence the global up-down symmetry remains unbroken in the finite 
volume ground state $\Pz(B)$ for any value of $B>0$.
When $m(B)>0$, we might say 
that the symmetry breaking in the infinite volume limit is 
``obscured'' by ``quantum fluctuation'' in finite systems.

A sign of the ``obscured symmetry breaking'' can be found 
as a long range order in the 
two-point correlation functions.
Although we have $(\Pz(B),S^{(3)}_x\,\Pz(B))=0$ for any $B>0$,
we expect (and can prove for sufficiently small $B$) that
\begin{equation}
(\Pz(B),S^{(3)}_xS^{(3)}_y\,\Pz(B)) \simeq m(B)^2
\label{LROIsing}
\end{equation}
holds for sufficiently large $|x-y|$.

Another sign of the ``obscured symmetry breaking'' can be found if 
we consider the first excited (eigen)state $\PL^{(1)}(B)$ of $\ham$ and 
its energy $E^{(1)}_\Lambda$.
When we have $m(B)>0$, we expect
\begin{equation}
\EL^{(1)}-\EL^{(0)}\approx \exp[-\tau(B) L^d]
\label{expdecay}
\end{equation}
holds as $L\toinf$ with a positive finite constant $\tau(B)$, 
where $\EL^{(0)}$ denotes the ground state energy.
The states $\{\PL^{(1)}(B)\}_\Lambda$ in this situation 
are typical examples of ``low-lying eigenstates''\footnote{
A beginner of exact diagonalization  
might identify $\PL^{(1)}(B)$ as a finite-size counterpart of 
an excited state in the infinite system.
As will become clear soon, this is totally misleading.
}.
See Definition~\ref{DEFlls}.

Now we shall discuss about states in 
the infinite system.
A ground state in the infinite system may be defined by the 
thermodynamic limit
\begin{equation}
\omega_B(A):=\lim_\TDL(\Pz(B),A\,\Pz(B)),
\label{omegaB}
\end{equation}
where $A$ is an arbitrary local operator ({\em i.e.}, a polynomial of 
spin operators), and $\Pz(B)$ is the unique ground state of $\ham$.
The limit is well-defined if one takes suitable subsequence of 
lattices.
(See Section~\ref{SecIG} and Appendix~\ref{APgs} for details.)
Since the finite volume ground state $\Pz(B)$ respects the global 
up-down symmetry, so does the infinite volume ground state 
$\omega_B(\cdots)$.
In particular we have
\begin{equation}
\omega_B(S^{(3)}_x)=0,
\label{NOSB0}
\end{equation}
for any $x$.

One might suspect from the above construction 
and the relation (\ref{NOSB0}) that, when the 
symmetry breaking field $B'$ is vanishing,
there is no 
symmetry breaking even in the infinite volume limit\footnote{
This is another possible confusion, 
to which we sometimes encounter.
}.
From a physical point of view, however, this conclusion is 
unnatural and misleading.
One should recall that
 there are many situations in nature where we do 
observe a symmetry breaking in the absence of explicit symmetry 
breaking fields\footnote{
\label{BEfn}
A typical example is antiferromagnetism, in which a staggered 
magnetic field plays the role of symmetry breaking field.
No mechanism can generate a real staggered magnetic field in an 
antiferromagnetic material.
A more drastic example is the Bose-Einstein condensation, where
the symmetry breaking field should create and annihilate particles!
({\em Note added in August 1997:}
This statement about Bose-Einstein condensation is wrong.
Because of the particle conservation law (in the whole universe),
 physically relevant states 
always have a fixed electron number.
(If we want to take into account fluctuation in 
the particle number, then we must
consider a mixed state.)
Thus a physically realistic states never break symmetry in this problem.)
}.
It is indeed possible to develop mathematically sensible definitions 
of infinite volume ground states which are capable of describing a 
symmetry breaking without symmetry breaking 
fields.
We discuss precise definitions in Section~\ref{SecIG} 
(Definition~\ref{DEFgs}) and 
Appendix~\ref{APgs}.
Here we shall see concrete examples.

Before discussing the symmetry breaking, however, 
let us observe that the above 
ground state $\omega_B(\cdots)$ indeed has an unnatural property.
Let $\Omega$ be a hypercubic region in ${\bf Z}^d$, and denote by 
$|\Omega|$ the number of sites in $\Omega$.
Consider the bulk physical quantity 
$M_\Omega :=\sum_{x\in\Omega}S^{(3)}_x$.
By combining (\ref{LROIsing}), (\ref{NOSB0}),
and Lemma~\ref{flucLemma1}, we find that
\begin{equation}
\frac{1}{|\Omega|^2}
\omega_B\sbk{\rbk{M_\Omega-\omega_B(M_\Omega)}^2}
\ge
m(B)^2,
\label{nocluster}
\label{cof}
\end{equation}
as $|\Omega|\toinf$.
The relation (\ref{nocluster}) implies that, in the state $\omega_B(\cdots)$,
the intensive bulk quantity $M_\Omega/|\Omega|$ has a finite fluctuation
provided that $m(B)>0$.
This is in contrast to the basic requirement in physics that any intensive 
bulk
quantity exhibits essentially no fluctuation in a thermodynamically stable
phase.
An infinite volume state in which any 
intensive bulk quantity has vanishing fluctuation
is called an ergodic state.
(See Definition~\ref{DEFergodic}.)
It is believed that a physically realizable state in a large system 
can be well approximated by an ergodic state.
(See Remark 1 at the end of the present section for further discussions.)
The behavior (\ref{nocluster}) implies that the state 
$\omega_B(\cdots)$ 
is not ergodic, and is hence unphysical.

Then there must be some physically natural ground states.
Let us note that the ground state $\Pz(B)$ and the first excited state 
$\PL^{(1)}(B)$ 
inherit the existence of symmetry breaking in the 
infinite volume limit.
We expect that, when $m(B)>0$, these states can be written as
\begin{equation}
\Pz(B)\simeq\frac{1}{\sqrt{2}}
\rbk{\tilde{\Phi}^+_\Lambda(B)+\tilde{\Phi}^-_\Lambda(B)},
\label{PL=}
\end{equation}
and
\begin{equation}
\PL^{(1)}(B)\simeq\frac{1}{\sqrt{2}}
\rbk{\tilde{\Phi}^+_\Lambda(B)-\tilde{\Phi}^-_\Lambda(B)},
\label{PL1=}
\end{equation}
where $\tilde{\Phi}^+_\Lambda$ and $\tilde{\Phi}^-_\Lambda$ are 
the states obtained by taking into account local quantum 
fluctuations into the completely ordered states 
$\PL^+$ and $\PL^-$, respectively.
Equations (\ref{PL=}) and (\ref{PL1=}) motivate us to 
define two states in the infinite system by
\begin{eqnarray}
\omega^\pm_B(A)&:=&\lim_\TDL
\frac{1}{2}\rbk{(\Pz(B)\pm\PL^{(1)}(B)),A
\,(\Pz(B)\pm\PL^{(1)}(B))},
\ret
&\simeq&\lim_\TDL
(\tilde{\Phi}^\pm_\Lambda(B),A\,\tilde{\Phi}^\pm_\Lambda(B)),
\label{omegapm}
\end{eqnarray}
for an arbitrary local operator $A$.
By using (\ref{expdecay}), the translation invariance of the expectation 
values, and the fact that
$(\PL^{(1)}(B),\ham\Pz(B))=0$, 
we find that these states satisfy
\begin{equation}
\omega^\pm_B(h_x)=\epsilon_0:=\omega_B(h_x),
\label{GS0}
\end{equation}
for any $x\in{\bf Z}^d$.
The local Hamiltonian is
$h_x=-\sum_{y;|x-y|=1}S^{(3)}_xS^{(3)}_y/2-BS^{(1)}_x$,
where the sum runs over the sites $y$ which are neighboring to $x$.
We call the above
$\epsilon_0$ the ground state energy density.
Following Definition~\ref{DEFgs}, 
we shall interpret the relation (\ref{GS0}) as indicating that the 
states $\omega^\pm_B(\cdots)$ are infinite volume ground 
states\footnote{
The existence of ground states other than $\omega_B(\cdots)$
apparently contradicts with the ``uniqueness of the ground
state'' we mentioned earlier, and has been a source of confusion
(especially in much more delicate situations, e.g., in Heisenberg
antiferromagnets).
Of course there is no contradiction, since the uniqueness (as is
proved by the Perron-Frobenius argument \cite{LiebMattis}) 
applies only to 
a finite system.
}.
We stress that this is a natural definition of ground states.
In a bulk (or an infinite) system, it is no longer meaningful to talk 
about small difference in the total energy.
What really count are the expectation values of the local 
Hamiltonian, and the present definition is designed precisely to 
look only at them.
We shall discuss more about the definitions of infinite volume 
ground states in Appendix~\ref{APgs}.

The final expression in (\ref{omegapm}) suggests 
the existence of an explicit symmetry breaking as
\begin{equation}
\omega_B^\pm(S^{(3)}_x)=\pm m(B).
\label{oMmM}
\end{equation}
If we assume the existence of a long range order as in 
(\ref{LROIsing}) and the existence of a gap
above $E_\Lambda^{(1)}$, we can prove the relation (\ref{oMmM})
from (\ref{OrderAP}) and Lemma~\ref{1=PsiLemma}.
Under the same assumptions, we can also prove that
the infinite volume ground state 
$\omega_B^\pm(\cdots)$ are ergodic.
See Theorem~\ref{ergodicth}.
We conclude that $\omega_B^\pm(\cdots)$
constructed by taking 
linear combinations of $\Pz(B)$ and $\PL^{(1)}(B)$
are the physically natural ground states in the infinite volume.

Let us summarize what we have learned from the 
present simple example.
When there is an ``obscured symmetry breaking'', we have the following.

\begin{itemize}
\item
There inevitably 
exists a ``low-lying eigenstate''.

\item
The infinite volume ground state defined by a naive infinite volume 
limit of finite volume ground states is not ``ergodic'', i.e., is 
unphysical.

\item
An ergodic ground state may be formed by taking a linear combination of 
the finite volume ground state and the ``low-lying (eigen)state'', and then 
taking the infinite volume limit.
\end{itemize}

In Section~\ref{Sec2}, we will see that these features are typical 
when there is an ``obscured symmetry breaking''.
We will mainly concentrate on
how the situation is modified when the relevant symmetry
is a continuous one.

\bigno
{\bf Remarks:}
1.
The statement that ``a physically realizable state in a large system 
can be well approximated by an ergodic state'' perhaps requires some
explanations.
Since this is a very delicate problem about observations in quantum many-body
systems, we can only give some heuristic ideas.

Consider a large but finite system at an extremely low temperature.
(Note that it is impossible to attain the absolute zero temperature
as long as observations are done within a finite amount of time.)
Suppose that the thermal energy is much larger than the excitation energy of the
``low-lying eigenstate'', which is quite likely since the thermal energy is proportional
to the system size.
Then one has a chance to 
observe not only the ground state but any linear combination of
the ground state and the ``low-lying state''.
Most of these linear combinations, however, suffer from the pathological
behavior that some bulk intensive quantities have finite fluctuations
as in (\ref{nocluster}).
A conventional wisdom suggests that a state with such a pathologically
large fluctuation is unstable under perturbations.
Small thermal disturbance may well destroy such a state, and bring it into
a more stable one.
We expect that such a mechanism will select only ergodic states out of the
infinitely many linear combinations of the ground state and the ``low-lying
state''.

\bigno
2. 
The reader might wonder about the nature of the ground state 
$\Pz(B)$ and the first eigenstate $\PL^{(1)}(B)$ when $B$ is large 
enough so that we have $m(B)=0$.
In this case, we expect that $\ham$ has a finite gap almost uniform 
in the lattice size $N$, and thus
\begin{equation}
\EL^{(1)}-\EL^{(0)}=O(1),
\end{equation}
as $N\toinf$.
According to the Definition~\ref{DEFlls}, 
the state $\PL^{(1)}(B)$ may be again called a ``low-lying 
eigenstate''.
However its nature is totally different from that in the case 
with $m(B)>0$.
(This may be regarded as 
a disadvantage of our definition of ``low-lying states''.
See the discussion following  Definition~\ref{DEFlls}.)

Roughly speaking, the first excited state $\PL^{(1)}(B)$ can be 
regarded 
as the state in which a single ``magnon'' is in the $k=0$ state, {\em 
i.e.\/},
\begin{equation}
\PL^{(1)}(B)\simeq\sum_{x\in\Lambda}\PL^{(x)}(B),
\label{magnon}
\end{equation}
where $\PL^{(x)}(B)$ is the state in which the magnon is localized at  
site $x$.
When $B$ is extremely large, $\PL^{(x)}(B)$ may be approximated by 
the 
state in which the spin at $x$ is pointing in the direction opposite to 
the magnetic field and all the other spins are pointing in the 
direction of the field.

The biggest difference from the case with $m(B)>0$ is that the 
limit
\begin{equation}
\tilde{\omega}_B(\cdots):=\lim_\TDL
((\alpha\Pz(B)+\beta\PL^{(1)}(B)),(\cdots)\,
(\alpha\Pz(B)+\beta\PL^{(1)}(B))),
\end{equation}
with any $\alpha$, $\beta$ with $|\alpha|^2+|\beta|^2=1$, defines 
exactly the same state as $\omega_B(\cdots)$ in (\ref{omegaB}).
More precisely, we have
\begin{equation}
\omega_B(A)=\tilde{\omega}_B(A),
\label{omega=tilomega}
\end{equation}
for an arbitrary local operator $A$.
The equality (\ref{omega=tilomega}) should be expected since, in an 
infinite system with only a single magnon, the probability of 
observing the magnon is vanishing.
We expect that, in this case, the infinite volume ground state is 
unique and preserves the global up-down symmetry.
Such a result can be proved rigorously for sufficiently large $B$.
See Theorem~\ref{UniqueGS}.

\Section{Results and physical consequences}
\label{Sec2}
In the present section, we describe our main results and their 
physical consequences in a general setting.
One of the goals is the construction of infinite volume ground states
with explicit symmetry breaking presented in Section~\ref{SecIG}.
A result on ``low-lying eigenstates'' which have direct relevance to
numerical approaches can be found in Section~\ref{SecFS}.
\subsection{Preliminaries}
\label{SecPre}
Here we fix some basic notations.
We also give a precise definition of ``low-lying states'',
and discuss motivations behind the definition.

We consider a quantum system 
on a finite lattice $\Lambda$ with $N$ sites. 
With each site $x \in \Lambda$, we associate a finite-dimensional 
Hilbert space ${\cal H}_x$. 
The full Hilbert space is
\begin{equation}
{\cal H}_\Lambda := \bigotimes_{x \in \Lambda} {\cal H}_x.
\label{Hilb}
\end{equation}
We note that a ``site'' in $\Lambda$ need not be an atomic site of a quantum
many-body system.
If necessary, one may call a group of atomic sites a ``site'', and
let $\calH_x$ be the corresponding finite dimensional Hilbert
space.

Throughout the present paper, the norm of a state 
$\PsL \in {\cal H}_\Lambda$ is defined as 
$\Vert \PsL \Vert ={(\PsL,\PsL)}^{1/2}$, 
and 
the norm of an operator $A$ on ${\cal H}_\Lambda$ as 
\begin{equation}
\Vert A \Vert := \sup_{\PsL \in {\cal H}_\Lambda}
{\Vert A\PsL \Vert \over \Vert \PsL \Vert}.
\label{DEFnorm}
\end{equation}

For a fixed $\Lambda$, we take the Hamiltonian 
\begin{equation}
\ham := \sum_{x \in \Lambda} \> h_x,
\label{ham}
\end{equation}
where each $h_x$ is a self-adjoint operator on ${\cal H}_\Lambda$. 

In order to discuss the notion of ``low-lying states'',
we take a sequence $\{\Lambda\}$ of finite lattices which tend to the 
infinite lattice ${\bf Z}^d$.
For each $\Lambda$ (with $N$ sites) we consider a quantum 
mechanical system on $\Lambda$ with the Hamiltonian 
$\ham$, and the ground state $\Phi^{(0)}_\Lambda$.
The corresponding eigenvalue of $\ham$ is denoted as $\EL^{(0)}$ .
\begin{definition}
A sequence of normalized
states $\{\PL'\}_\Lambda$ are called ``low-lying 
states'', if 
\begin{equation}
\lim_\TDL\frac{1}{N}\rbk{(\PL',\ham\,\PL')-\EL^{(0)}}=0
\label{LLS}
\end{equation}
holds, and if each $\PL'$ is orthogonal to the ground state $\PL^{(0)}$.
``Low-lying states'' in which each state $\PL'$ happens to be 
an eigenstate of the 
Hamiltonian $\ham$ are called ``low-lying eigenstates''.
\label{DEFlls}
\end{definition}

The above definition of ``low-lying states'' is mainly motivated by 
Theorem~\ref{LLS=GS}, which says that any translation invariant 
``low-lying states'' converge to an infinite volume ground state.
We note, however, that the definition is too general to
indicate only those ``low-lying states'' which play crucial roles in
forming infinite volume ground states with symmetry breaking.
For example, the ``magnon state'' (\ref{magnon}) discussed in the
remark of section~\ref{SecEx} is also a ``low-lying state'' according to the
definition, but one usually wishes to consider it as an excited
state\footnote{
It is worth mentioning, however, that the state with exactly one
magnon is never observed as an excited state in actual experiments.
One can measure the effects caused by magnons only in the state
where magnons have a finite density.
Such a state is, of course, not a ``low-lying state''.
}.
The reader may regard that the definition is introduced for notational
convenience, rather than to indicate a physically important notion\footnote{
If we recall the discussion in Remark 1 of Section~\ref{SecEx}, however,
it is possible to give a physical meaning to the above definition.
The equation (\ref{LLS}) precisely states the condition that the energy
gap of the states $\{\PL'\}$ is dominated by the thermal energy
in a sufficiently large system.
}.

\subsection{Theorem of Horsch and von der Linden}
\label{SecHV}
Before discussing our own results, we describe the theorem due to 
Horsch and von der Linden\cite{HorschLinden}, 
which is the first rigorous result 
concerning the existence of a ``low-lying state'' in the presence of an 
``obscured symmetry breaking''.

We consider a finite lattice $\Lambda$ with $N$ sites, and a quantum
many-body system on it as in Section~\ref{SecPre}.
Let
\begin{equation}
\op:=\sum_{x\in\Lambda}o_x
\label{ord}
\end{equation}
be the order operator, where each $o_x$ is a self-adjoint operator 
on ${\cal H}_\Lambda$.

Assume that  $\norm{h_x}\le h$ and $\norm{o_x}\le o$ hold 
for any $x$
with $x$-independent finite constants $h$ and $o$.
Assume also that
$[o_x,o_y]=0$ holds for any $x,y$, and $[h_x,o_y]=0$ 
holds unless the site $y$ belongs to the support set ${\cal S}_x$.
We require that the number of sites in 
${\cal S}_x$ is bounded from above by an integer $r$.
Let $\PL$ be an eigenstate of $\ham$ with the eigenvalue $\EL$.
We assume that the state $\PL$ exhibits an ``obscured symmetry 
breaking'' in the sense that it satisfies
\begin{equation}
\bkt{\op}=0,
\label{NOorder0}
\end{equation}
and
\begin{equation}
\bkt{(\op)^2}\ge (\mu oN)^2
\label{LRO0}
\end{equation}
with a constant $0<\mu\le1$.
We define
\begin{equation}
\PsL:=\frac{\op\PL}{\norm{\op\PL}},
\end{equation}
which is well-defined since $\norm{\op\PL}$ is 
nonvanishing because of (\ref{LRO0}).

Then the theorem of Horsch and von der Linden is the following.
If we set $\PL$ as the ground state $\Pz$,
it shows that there is a sequence of states $\{\PsL^{(1)}\}_\Lambda$ 
which form ``low-lying 
eigenstates'' (in the sense of Definition~\ref{DEFlls}), and their excitation
energy is less than of order $N^{-1}$.
\begin{theorem}
The expectation value of the energy in the state $\PsL$ satisfies
\begin{equation}
\frac{1}{N}\abs{(\PsL,\ham\,\PsL)
-\EL}
\le c_0\frac{1}{N^2}
\end{equation}
with $c_0=2r^2h\mu^{-2}$.
When $\PL$ is the ground state $\Pz$ of $\ham$, there exists an 
eigenstate $\PL^{(1)}$ of $\ham$ whose energy $E^{(1)}_\Lambda$ satisfies
\begin{equation}
E^{(1)}_\Lambda-\EL\le c_0\frac{1}{N}.
\end{equation}
\label{M=1theorem}
\end{theorem}
\begin{proof}{Proof}
A crucial observation is that the energy difference can be expressed
in terms a double commutator as the following.
Then the first part follows as
\begin{eqnarray}
\abs{(\PsL,\ham\,\PsL)
-\EL}
&=&
\frac{\abs{\bkt{\sbk{\sbk{\op,\ham},\op}}}}{2\norm{\op\PL}^2}
\ret
&\le&\frac{\norm{\sbk{\sbk{\op,\ham},\op}}}{2\norm{\op\PL}^2}
\ret
&\le&
\frac{4r^2ho^2N}{2(\mu oN)^2}=c_0\frac{1}{N},
\label{dc}
\end{eqnarray}
where we have used the assumed commutation relations
and the norm bounds of $h_x$ 
and $o_x$, as well as (\ref{LRO0}).
To prove the second part, we note that the relation 
(\ref{NOorder0}) implies that the state $\PsL$ is orthogonal to the 
ground state $\Pz$.
Then the statement in the theorem is a simple consequence of the 
variational principle.%
\end{proof}

\subsection{Main theorems}
\label{SecMain}
Now we shall describe our own theorems about ``low-lying states''.
They apply to a system with a continuous symmetry, and 
establish the existence of ever increasing numbers of ``low-lying 
eigenstates''.

We again consider the finite lattice $\Lambda$ with $N$ sites, 
and a quantum many-body system on it as in Section~\ref{SecPre}.
We further require that the system possesses a global $U(1)$ 
symmetry, 
whose generator is a self-adjoint operator $\gen$.
We assume that 
\begin{equation}
\sbk{\ham,\gen}=0.
\end{equation}
We introduce the order operators 
\begin{equation}
O_\Lambda^{(\alpha)}:=\sum_{x\in\Lambda}o_x^{(\alpha)},
\label{orderOp}
\end{equation}
where $\alpha=1,2$, and each $o_x^{(\alpha)}$ is a self-adjoint 
operator on ${\cal H}_\Lambda$.
The order operators form two components of a vector 
$(\opone,\optwo)$ which transforms under the action of $U(1)$, 
and measure a possible spontaneous breakdown of 
the $U(1)$ symmetry.
They satisfy the standard commutation relations
\begin{equation}
\sbk{\opone,\gen}=-i\optwo,\quad
\sbk{\optwo,\gen}=i\opone.
\label{commutation1}
\end{equation}
We also introduce
\begin{equation}
O^\pm_\Lambda:=\opone\pm i\optwo,
\label{Oplusminus}
\end{equation}
which satisfy the commutation relations
\begin{equation}
\sbk{\opplus,\gen}=-\opplus,\quad
\sbk{\opminus,\gen}=\opminus.
\label{commutation2}
\end{equation}
The operators $\opplus$ and $\opminus$ are the raising 
and the lowering operators, respectively, for the quantum number 
defined by the self-adjoint operator $\gen$.

We assume that these operators satisfy the following 
three conditions. 
\par\bigskip
\noindent
i) \quad
$[o_x^{(\alpha)},o_y^{(\beta)}]=0$ for $x\ne y$ and 
$\alpha,\beta=1,2$.
\par\bigskip
\noindent
ii) \quad
$ [h_x, o_y^{(\alpha)}] = 0 $ 
holds for $\alpha=1,2$ unless $ y \in{\cal S}_x$.
The number of sites in the support set 
${\cal S}_x \subset \Lambda$ is 
bounded from above by an $x$-independent integer $r\geq2$.
\par\bigskip
\noindent
iii) \quad
There are $x$-independent finite constants $h$ and  $o$, and we have 
$\Vert h_x \Vert \le h$
and 
$\Vert o_x^{(\alpha)} \Vert \leq o$
for any $x \in \Lambda$ and $\alpha=1,2$. 

\par\bigskip
Let  $\PL$  be a normalized simultaneous eigenstate 
of the Hamiltonian $\ham$ and the self-adjoint operator 
$C_\Lambda$.
We denote by $\EL$ the corresponding eigenvalue of $\ham$.
Usually we take $\PL$ as the ground state $\Pz$ of $\ham$.
We assume that 

\par\bigskip
\noindent
iv) \quad
the state $\Phi_\Lambda$ exhibits a long range order 
in the sense that 
\begin{equation}
\bkt{(\opone)^2}=\bkt{(\optwo)^2}
\geq(\mu o N)^2
\label{LRO}
\end{equation}
holds with a constant $0<\mu\le1$.

\par\bigskip
From the commutation relations (\ref{commutation2}) and the fact 
that $\PL$ is an eigenstate of $\gen$, we automatically have
\begin{equation}
\bkt{\opone}=\bkt{\optwo}=0.
\label{NOSB}
\end{equation}
In other words, the state $\PL$ have vanishing order parameters.
The relations (\ref{LRO}) and (\ref{NOSB}) together imply that the 
state 
$\PL$ exhibits an ``obscured symmetry breaking''\footnote{
In fact  (\ref{LRO}) and (\ref{NOSB}) may well hold in a finite system
whose infinite volume limit does not exhibit a symmetry breaking,
in which case the parameter $\mu$ vanishes as $\TDL$.
What we really mean by an ``obscured symmetry breaking'' is
that  (\ref{LRO}) and (\ref{NOSB}) are valid with a 
$\Lambda$-independent $\mu>0$.
}.

For a nonvanishing integer $M$, we consider the state
\begin{equation}
\PsL^{(M)}:=\frac{(\opplus)^M\PL}
{\norm{(\opplus)^M\PL}},
\label{Psi}
\end{equation}
where we set $(\opplus)^M=(\opminus)^{-M}$ for a negative $M$.
Although the state (\ref{Psi}) is ill-defined if 
$(\opplus)^M\Phi_\Lambda =0$,
the following theorems guarantee that this is not the case when 
certain conditions are met.

The first theorem of the present paper is the following.
Although the bound (\ref{B2}) for the energy expectation value 
may not look quite strong, it will turn out to be sufficient for 
a construction of infinite volume ground states with symmetry breaking.
A better estimate for the energy expectation value will be provided in the
second theorem.
\begin{theorem}
When the assumptions~i)-iv) are valid, and we further have
\begin{equation}
N\ge \rbk{\frac{4r}{\mu}}^2,
\label{N>}
\end{equation}
and
\begin{equation}
\frac{\abs{M}}{N}\le\frac{\mu^2}{8r},
\label{M<}
\end{equation}
the state $\PsL^{(M)}$ of (\ref{Psi}) is well-defined.
The expectation value of the 
energy in the state satisfies 
\begin{equation}
 \frac{1}{N}\left|(\PsL^{(M)} , \ham 
\PsL^{(M)}) 
 - \EL\right|
\le c_1\frac{|M|}{N},
\label{B2}
\end{equation}
where $c_1$ is a constant which depends only on $h$, $r$, and 
$\mu$.
\label{generalMtheorem}
\end{theorem}

By taking $\PL$ as the ground state $\PL^{(0)}$ of the Hamiltonian,
the theorem implies that $\{\PsL^{(M)}\}_\Lambda$ with a fixed $M$
form ``low-lying states'' in the sense of Definition~\ref{DEFlls}.
The theorem will be proved in Section~\ref{Sec4}.

In order to get a better estimate of the energy difference (at least 
for small enough $|M|$), we require higher symmetry.
\par\bigno
v)\quad
The order operators satisfy the commutation relation
\begin{equation}
\sbk{\opone,\optwo}=i\gamma\gen,
\label{commutation3}
\end{equation}
where $\gamma$ is a real constant.
\par\bigskip
In other words, we assume that the vector 
$(\gamma^{-1/2}\opone,\gamma^{-1/2}\optwo,\gen)$ form 
generators of $SU(2)$.
We do not, however, assume that the system has a full $SU(2)$ 
symmetry.
We only require a partial $U(1)\times{\bf Z}_2(\cong O(2))$ symmetry as 
follows.
\par\bigskip
\noindent
vi)\quad
We have
$U_\Lambda H (U_\Lambda)^{-1}=H$ and $U_\Lambda\PL\propto\PL$,
where $U_\Lambda=\exp[i(\pi/\sqrt{\gamma})\opone]$ represents 
the $\pi$-rotation around the first axis.
\par\bigskip
Then the second theorem is as follows.
\begin{theorem}
When the assumptions i)-vi) are valid, and we further have
\begin{equation}
\frac{M^2}{N}\le c_2,
\label{M/N<}
\end{equation}
with $c_2=\min\{\mu^2/(192r),o\mu/\sqrt{24\gamma}\}$,
the state $\PsL^{(M)}$ of (\ref{Psi}) is well-defined.
The expectation value of the 
energy in the state satisfies 
\begin{equation}
 \frac{1}{N}\left|(\PsL^{(M)} , \ham 
\PsL^{(M)}) 
 - \EL\right|
\le c_3\rbk{\frac{M}{N}}^2,
\end{equation}
where $c_3$ is a constant which depends only on 
$h$, $o$, $r$, $\mu$, 
and $\gamma$.
\label{smallMtheorem}
\end{theorem}

This theorem has a rather strong implication on the property of
the low-lying spectrum of the Hamiltonian $\ham$.
See Section~\ref{SecFS}.
The theorem will be proved in Section~\ref{Sec5}.

\bigskip\noindent
{\bf Remark:}
The condition vi) for Theorem~\ref{smallMtheorem} can be weakened 
to
$\gen\PL=0$ if one replaces the definition of the trial state 
(\ref{Psi}) by
\begin{equation}
\PsL^{(M)}:=\frac{(\opplus)^M\PL+(\opminus)^M\PL}
{\norm{(\opplus)^M\PL+(\opminus)^M\PL}}.
\end{equation}
\subsection{States with explicit symmetry breaking}
As we have already mentioned in the introduction,
the particular set of states $\{\PsL^{(M)}\}$ defined in (\ref{Psi})
is not introduced as mere candidates of ``low-lying states''.
These states have a special feature that they
can be regarded as ``parts'' of infinite volume ground
states with explicit symmetry breaking.
In order to demonstrate this fact, we shall here construct a sequence of
``low-lying states'' which exhibits symmetry breaking.
The basic idea in the present construction appeared already in
our earlier publication \cite{KomaTasaki}.

Let us consider a sequence of models which satisfy the conditions 
i)-iv) of Section~\ref{SecMain} with the state $\PL$ chosen as the
ground state $\PL^{(0)}$ of the Hamiltonian $\ham$.
Note that we are only assuming the $U(1)$ symmetry as is required for
Theorem~\ref{generalMtheorem}.
For a positive integer $k$, we take a linear combination of the ground
state and the ``low-lying states'' $\PsL^{(M)}$ as
\begin{equation}
\Xi^{(k)}_\Lambda:=\frac{1}{\sqrt{2k+1}}
\cbk{\Pz+\sum_{M=1}^k
\rbk{\PsL^{(M)}+\PsL^{(-M)}}}.
\label{Xidef}
\end{equation}
We shall take the lattice $\Lambda$ sufficiently large so that the bounds 
(\ref{N>}) and (\ref{M<}) are valid for any $M$ with $|M|\le k$.
By using Theorem~\ref{generalMtheorem}
and the fact that $(\PsL^{(i)},\ham\PsL^{(j)})=0$ for $i\ne j$,
we note that the state $\Xi^{(k)}_\Lambda$ is normalized,
and satisfies
\begin{equation}
\lim_\TDL \frac{1}{N}
\cbk{(\Xi^{(k)}_\Lambda,\ham\,\Xi^{(k)}_\Lambda)-
\EL}=0,
\label{XihasE=0}
\end{equation}
for any $k$.
Thus $\{\Xi^{(k)}_\Lambda\}_\Lambda$ with a fixed $k$ form 
``low-lying states''.

A remarkable properties of these ``low-lying states'' are the following.
\begin{theorem}
The expectation value of the order operators in the state 
$\Xi^{(k)}_\Lambda$ satisfy
\begin{equation}
(\Xi^{(k)}_\Lambda,\optwo\,\Xi^{(k)}_\Lambda)=0,
\label{O2=0}
\end{equation}
and
\begin{equation}
\lim_{k\toinf}\lim_\TDL \frac{1}{N}
(\Xi^{(k)}_\Lambda,\opone\,\Xi^{(k)}_\Lambda)
\ge\sqrt{2}\mu o,
\label{XiOXi}
\end{equation}
where the prefactor $\sqrt{2}$ is modified if the model has a 
higher symmetry.
For example, we replace $\sqrt{2}$
with $\sqrt{3}$ 
when 
the model has an $SU(2)$ symmetry.
\label{KTtheorem}
\end{theorem}
\begin{proof}{Outline of proof}
We first note that
\begin{eqnarray}
(\Xi^{(k)}_\Lambda,\opplus\,\Xi^{(k)}_\Lambda) &=&
\frac{1}{2k+1}\sum_{M=-k}^k\sum_{M'=-k}^k
\frac{\bktz{(\opplus)^{-M}\opplus(\opplus)^{M'}}}
{\norm{(\opplus)^M\Pz}\norm{(\opplus)^{M'}\Pz}}
\ret
&=&
\frac{1}{2k+1}\sum_{M=-k+1}^k
\frac{\bktz{(\opplus)^{-M}\opplus(\opplus)^{M-1}}}
{\norm{(\opplus)^M\Pz}\norm{(\opplus)^{M-1}\Pz}}
\ret
&=&
\frac{1}{2k+1}\sum_{M=1}^k
\frac{\bktz{(\opminus)^{M}(\opplus)^{M}}}
{\norm{(\opplus)^M\Pz}\norm{(\opplus)^{M-1}\Pz}}
\ret
&&+
\frac{1}{2k+1}\sum_{M''=1}^k
\frac{\bktz{(\opplus)^{M''}(\opminus)^{M''}}}
{\norm{(\opminus)^{M''-1}\Pz}\norm{(\opminus)^{M''}\Pz}},
\end{eqnarray}
where we used the shorthand notation $(\opplus)^{-M}=(\opminus)^M$
for $M>0$.
(It must be noted that $\opminus$ is not the inverse of $\opplus$.)
We have used the facts that $\PL$ is an eigenstate of $\gen$
and $\op^\pm$ are the raising and lowering operators for the charge
defined by $\gen$ to get the second equality.
To get the final line, we have set $M''=1-M$.
A similar calculation for $(\Xi^{(k)}_\Lambda,\opminus\,\Xi^{(k)}_\Lambda)$
shows that 
$(\Xi^{(k)}_\Lambda,\opplus\,\Xi^{(k)}_\Lambda)
=(\Xi^{(k)}_\Lambda,\opminus\,\Xi^{(k)}_\Lambda)$, and
hence the desired relation (\ref{O2=0}).
Note that we do not have to use the ${\bf Z}_2$ symmetry as is assumed
in the conditions v), vi) of Section~\ref{SecMain}.

The relation (\ref{XiOXi}) is essentially proved in \cite{KomaTasaki}.
One only has to combine  (7.26) of \cite{KomaTasaki} and Theorem~6.1
of \cite{KomaTasaki}. 
Although some estimates in \cite{KomaTasaki} implicitly assume the
larger $U(1)\times{\bf Z}_2$ symmetry, this is not necessary.
A careful treatment (as we did above) shows that all the estimates
in \cite{KomaTasaki} are valid for the models with only a global
$U(1)$ symmetry\footnote{
One can considerably improve the estimates in \cite{KomaTasaki}
by using the techniques developed in Section~\ref{Sec5} of the present
paper.
}.%
\end{proof}

The theorem establishes that the state $\Xi^{(k)}_\Lambda$ 
exhibits 
explicit symmetry breaking.
By applying the $U(1)$ rotation $\exp[i\theta\gen]$ to the state 
$\Xi^{(k)}_\Lambda$, we also get states in which the order 
parameter is pointing different directions.

\subsection{Infinite volume ground states}
\label{SecIG}
Now we shall discuss the relation between the ``low-lying 
states'' in the sequence of finite systems and the 
infinite volume ground states.
We again observe that, when there is 
an  ``obscured symmetry breaking'', 
the naive infinite volume limit of the finite 
volume ground states is not an ergodic state, and is hence unphysical.
By forming suitable linear combinations of the (finite-volume) 
ground state and the
``low-lying states'', and then taking  infinite 
volume limits, we get infinite volume ground states
with explicit symmetry breaking.
We conjecture these infinite volume ground states to be ergodic,
{\em i.e.\/}, physically natural.

In order to simplify the discussion, we make several assumptions 
on 
the model.
We assume that each finite lattice $\Lambda$ is a $d$-dimensional 
hypercubic lattice with periodic boundary conditions.
We again denote by $N$ the number of sites in $\Lambda$.
We assume that the Hamiltonian (\ref{ham}) 
and the order operators (\ref{orderOp}) are translation 
invariant 
in the sense that we can write
$h_x=\tau_x(h_o)$ and $o^{(\alpha)}_x=\tau_x(o^{(\alpha)}_o)$ 
for any $x$.
Here $\tau_x$ is the translation by the lattice
vector $x$ (which respects the periodic boundary conditions),
and the operators $h_o$ and $o_o$
are independent of $\Lambda$.

A local operator $A$ is an operator which acts nontrivially only on 
a finite number of sites (or, more precisely, on a finite dimensional 
Hilbert space $\bigotimes_{x\in{\cal S}(A)}{\cal H}_x$ with a finite 
support set ${\cal S}(A)$).
Let 
$\rho_\Lambda(\cdots)=
{\rm Tr}_{\calH_\Lambda}[(\cdots)\widetilde{\rho}_\Lambda]$ 
be a state of the system on $\Lambda$, where 
$\widetilde{\rho}_\Lambda$ is an arbitrary density matrix
on $\calH_\Lambda$.
Given a sequence of (finite-volume) states 
$\{\rho_\Lambda(\cdots)\}_\Lambda$, we (formally) define
\begin{equation}
\rho(A):=\lim_\TDL \rho_\Lambda(A)
\label{rholim}
\end{equation}
for each local operator $A$.
The above $\rho(\cdots)$ is a linear map from the space of 
local 
operators to the set of complex numbers ${\bf C}$.
We call $\rho(\cdots)$ a {\em state\/} of the infinite system.
(See Appendix~\ref{APgs} for the general definition of a state in an infinite 
system.)
It might happen, however, that the limit (\ref{rholim})
does not exist for all local 
$A$.
It is known that one can always choose a subsequence of 
lattices so 
that the limit is well-defined.
See Appendix~\ref{APgs} for a proof.
(An elementary proof can be constructed by using the
``diagonal sequence trick'', as is 
illustrated, e.g., in Theorem~I.24 of \cite{ReedSimon}.)

We want to describe what we mean by ground states of the infinite 
system.
Since it is meaningless to talk about eigenstates or eigenvalues of 
the total Hamiltonian $\ham$ when $\TDL$, a different point of 
view is necessary.
Here we employ probably the simplest definition for ground 
states of an infinite system.
As we discuss in Appendix~\ref{APgs}, the present definition is 
equivalent to the other definitions which are standard in 
mathematical literature.
It simply says that a ground state should 
minimize the 
local energy.

\begin{definition}
We define the ground state energy density $\epsilon_0$ by
\begin{equation}
\epsilon_0:=\lim_\TDL
\inf_{\begin{array}{l}
{}^{\Phi_\Lambda\in{\cal H}_\Lambda}_{\norm{\Phi_\Lambda}=1}
\end{array}
}\frac{1}{N}
\bkt{\ham},
\label{epsilon0}
\end{equation}
where the limit always exists.
An infinite volume state $\omega(\cdots)$ is said to be a  
ground state if it satisfies
\begin{equation}
\omega(h_x)=\epsilon_0,
\label{GS}
\end{equation}
for any $x\in{\bf Z}^d$.
\label{DEFgs}
\end{definition}

We also introduce the precise notion of ergodic states in an
infinite system.
In the following, we shall give a simple intuitive definition.
See the remark at the end of the present section for 
the relation between the present definition and other
related notions.

In short the definition says that a state is ergodic 
if and only if any intensive 
bulk quantity
has essentially no fluctuation in the state.
Since the requirement is believed to apply to any physically realizable
state of a large system, we might say that a translation invariant
state is physically natural if and only if it is ergodic.
(See Remark 1 of Section~\ref{SecEx}.)
It is also known that a non-ergodic translation invariant state can be
decomposed into an ``integral'' over ergodic states.
See the remark at the end of the present section.

\begin{definition}
Let $\Omega$ be a hypercubic region in ${\bf Z}^d$, and denote the number of
sites in $\Omega$ by $|\Omega|$.
For an arbitrary local self-adjoint operator $A$, we define the corresponding
bulk quantity as 
$A_\Omega:=\sum_{x\in\Omega}\tau_x(A)$,
where $\tau_x(A)$ is the translate of $A$ by a lattice vector $x$.
Let $\rho(\cdots)$ be a translation invariant state, {\em i.e.\/}, a state
which satisfies $\rho(B)=\rho(\tau_x(B))$ for any local
operator $B$ and any $x\in{\bf Z}^d$.
The state $\rho(\cdots)$ is said to be ergodic
if, for any $A$, the intensive bulk quantity $A_\Omega/|\Omega|$
exhibits vanishing fluctuation in the sense that
\begin{equation}
\lim_{|\Omega|\toinf}\frac{1}{|\Omega|^2}
\rho\sbk{\rbk{A_\Omega-\rho(A_\Omega)}^2}=0.
\end{equation}
\label{DEFergodic}
\end{definition}

For each finite $\Lambda$, let $\Pz$ be a ground state of 
$\ham$.
We can assume $\Pz$ is translation invariant since the Hamiltonian
is.
Then it is easy to verify that the infinite volume state defined by
\begin{equation}
\omega(A):=\lim_\TDL \bktz{A},
\label{infGSseq}
\end{equation}
for any local operator $A$ (by taking a suitable subsequence), 
is indeed an infinite-volume ground state in the sense of
Definition~\ref{DEFgs}.

Assume that each finite volume ground state $\Pz$ exhibits 
an ``obscured 
symmetry breaking'' in the sense that it satisfies (\ref{LRO}) and 
(\ref{NOSB}).
Then by using Lemma~\ref{flucLemma1}, we can show that
\begin{equation}
\frac{1}{|\Omega|^2}\omega
\rbk{(O^{(\alpha)}_\Omega-\omega(O^{(\alpha)}_\Omega))^2}
\ge (\mu o)^2,
\label{omegaNOcluster}
\end{equation}
for any finite region $\Omega\subset{\bf Z}^d$,
where $O^{(\alpha)}_\Omega=\sum_{x\in\Omega}o^{(\alpha)}_x$.
This implies that the ground state $\omega(\cdots)$
is not an ergodic state, and is hence unphysical.

We still do not know how to construct ergodic ground states in 
a general setting.
In Appendix~\ref{APergodic}, however, we present a general
construction of ergodic
infinite volume ground states in a system where a {\em discrete}
symmetry is spontaneously broken and a gap above the first
``low-lying eigenstate'' is generated.
In what follows, we make some observations which
suggest that the similar construction as in 
Appendix~\ref{APergodic} might work in systems with a
broken continuous symmetry.

Let us start from a simple but important theorem
which summarizes the
relation between ``low-lying states'' and ground states of an 
infinite system.
\begin{theorem}
Let $\{\PL'\}_\Lambda$ be ``low-lying states'' in the sense of
Definition~\ref{DEFlls},
and assume that each $\PL'$ defines 
translation invariant expectation values, i.e.,
$(\PL',A\,\PL')=(\PL',\tau_x(A)\,\PL')$ for any
$x\in\Lambda$ and for any local operator $A$.
Then the state
\begin{equation}
\omega'(\cdots):=\lim_\TDL(\PL',(\cdots)\,\PL'),
\end{equation}
defined by taking a suitable subsequence of lattices, is a ground 
state.
\label{LLS=GS}
\end{theorem}
\begin{proof}{Proof}
The translation invariance implies
\begin{equation}
(\PL',\ham\,\PL')=N(\PL',h_x\,\PL'),
\end{equation}
for any $x\in\Lambda$.
Then the condition (\ref{LLS}) of ``low-lying states'' reads
\begin{equation}
\lim_\TDL\cbk{(\PL',h_x\,\PL')-\bktz{h_x}}=0,
\end{equation}
which reduces to $\omega'(h_x)=\epsilon_0$ for any $x$.%
\end{proof}
It should be stressed that, in the above, the ``low-lying states'' $\PL'$ 
need 
not be ground states or eigenstates of finite systems.

The states $\Xi_\Lambda^{(k)}$ defined in (\ref{Xidef}), and 
its $U(1)$ rotations are ``low-lying states''
with translation invariant expectation values.
By using Theorem~\ref{LLS=GS} and Theorem~\ref{KTtheorem}, we
get the following important result which completes a construction of
infinite volume ground states with explicit symmetry breaking.
\begin{coro}
For $0\le\theta<2\pi$, define infinite volume states by
\begin{equation}
\omega_\theta(\cdots):=\lim_{k\toinf}\lim_\TDL
(e^{i\theta\gen}\,\Xi^{(k)}_\Lambda,(\cdots)\,
e^{i\theta\gen}\,\Xi^{(k)}_\Lambda),
\label{ergodic}
\end{equation}
where we take subsequences if necessary.
The states $\omega_\theta(\cdots)$ are infinite volume ground states.
They exhibit explicit symmetry breaking as
\begin{equation}
\omega_\theta\sbk{o_x^{(1)}}=m\cos\theta,\quad
\omega_\theta\sbk{o_x^{(2)}}=m\sin\theta,
\label{ergodicorder}
\end{equation}
for any $x$.
The order parameter $m$ satisfies
\begin{equation}
 m\ge\sqrt{2}o\mu,
 \label{m>sqr2}
\end{equation}
for systems
with a $U(1)$ symmetry, and $m\ge\sqrt{3}o\mu$ for systems with an 
$SU(2)$ symmetry.
\label{omegaCoro}
\end{coro}

It is believed that, in a system where a $U(1)$ symmetry 
is spontaneously broken,
the non-ergodic ground state $\omega(\cdots)$ 
(defined in (\ref{infGSseq})) is decomposed as
\begin{equation}
\omega(\cdots)=\frac{1}{2\pi}\int_0^\theta d\theta\,
\omega^{\rm ergodic}_\theta(\cdots),
\label{ergodicdeomp}
\end{equation}
where $\omega^{\rm ergodic}_\theta(\cdots)$ is an ergodic ground state
which satisfies 
\begin{equation}
\omega^{\rm ergodic}_\theta\sbk{o_x^{(1)}}=\mmax\cos\theta,\quad
\omega^{\rm ergodic}_\theta\sbk{o_x^{(2)}}=\mmax\sin\theta,
\end{equation}
where $\mmax>0$ is the maximum possible value of the order parameter
within the infinite volume ground states.

Let us examine how the order parameter $\mmax$ is related to the 
long range order observed in two point functions.
There are two different ways of defining the long range order parameter.
The first definition deals directly with the infinite volume state.
We define the long range order parameter $\mu_1$ for the (non-ergodic)
state $\omega(\cdots)$ by
\begin{equation}
\mu_1:=\lim_{\Omega\uparrow{\bf Z}^d}
\frac{1}{o|\Omega|}
\sqrt{\omega((O^{(1)}_\Omega)^2)}.
\label{mu1}
\end{equation}
The other definition deals with a sequence of finite volume ground state.
We define the long range order parameter $\mu_2$ as 
\begin{equation}
 \mu_2 := \limsup_\TDL\frac{1}{oN}\sqrt{\bktz{(\opone)^2}},
 \label{mu2}
\end{equation}
where $\Pz$ is the ground state on $\Lambda$.
Note that this definition is motivated by (\ref{LRO}), one of the basic
assumptions of the present paper.

It is quite likely that the above two definitions give the same
result for a large class of systems.
Unfortunately, we are only able to prove the one sided inequality
$\mu_1\ge\mu_2$.
(This follows from  Lemma~\ref{flucLemma1}.)
Let us proceed by assuming that the equality $\mu_1=\mu_2$
is valid.

Assuming the decomposition (\ref{ergodicdeomp}), we see,
for sufficiently large hypercubic region $\Omega$, that
\begin{equation}
\frac{1}{|\Omega|^2}\omega((O^{(1)}_\Omega)^2)
\simeq\frac{1}{2\pi}\int_0^\theta d\theta\,(\mmax\cos\theta)^2
=\frac{(\mmax)^2}{2},
\label{sqrt2}
\end{equation}
where we have used that $\omega^{\rm ergodic}_\theta(\cdots)$
is ergodic.
Then by using (\ref{mu1}) and (\ref{sqrt2}), we observe that 
$\mmax=\sqrt{2}o\mu_1$.
On the other hand, the inequality (\ref{m>sqr2}) and the definition
(\ref{mu2}) of the long range order parameter immediately implies
that $m\ge\sqrt{2}o\mu_2$.
Combining these two equations with the conjectured $\mu_1=\mu_2$,
we get $m\ge\mmax$.
Since $\mmax$ is defined as the maximum value of the order parameter,
this leads us to the (plausible but nonrigorous) 
conclusion that we indeed have $m=\mmax$.

This observation motivates us to state the following conjecture.
\begin{conjecture}
The infinite volume ground states 
$\omega_\theta(\cdots)$ defined in (\ref{ergodic})
are nothing but the desired ergodic (i.e., physically natural)
ground states $\omega^{\rm ergodic}_\theta(\cdots)$.
\end{conjecture}

Unfortunately we have no direct evidences which support the 
conjecture.
For systems with a discrete symmetry, however, we can prove that
the statement corresponding to the above conjecture is in fact valid.
See Appendix~\ref{APergodic}.

\bigskip\noindent
{\bf Remarks:}
1. Our Definition~\ref{DEFergodic} of ergodic states is actually not
exactly the same as the standard one \cite{BratteliRobinson,Ruelle,Simon}.
For the particular class of systems we are considering here, however, 
it turns out that our definition is equivalent to the standard definition of
${\bf Z}^d$-ergodic states.
See, for example, Sections~6.3 and 6.5 of \cite{Ruelle}.
(In fact, our definition is motivated by Lemma 6.5.1 of \cite{Ruelle}.)

There is a beautiful decomposition theory for ${\bf Z}^d$-ergodic states.
It states that an arbitrary non-ergodic translation invariant state
can be decomposed into a kind of ``integral'' over ergodic states.
See Section~I.7 (especially Theorem I.7.10) of \cite{Simon} 
or Section~6.4 of \cite{Ruelle}.
A detailed treatment of the decomposition theory can be found in
Chapter 4 of \cite{BratteliRobinson}.

A disadvantage in the notion of ergodic states (as in 
Definition~\ref{DEFergodic}) is that one has to
a priori assume the correct invariance of the states 
in order to get a physically natural states.
Another way of characterizing physically natural states is to make use of
the notion of {\em pure\/} states.
This notion is in some sense more abstract than that of ergodic states, and
does not make use of any 
specific invariance.
See \cite{BratteliRobinson,Ruelle} for the definition of
{\em pure} states and for what are known about it.
(One should be aware that the notion of pure states for an infinite system is
distinct from that for ordinary quantum mechanics with finite degrees of
freedom.)

\bigno
2.
Another way of (formally) defining infinite volume
ground states with explicit symmetry breaking is to apply
infinitesimal symmetry breaking field to the system.
Let $\Pz(B)$ be a ground state of the Hamiltonian
$\ham-B\opone$, where $B$ is real valued symmetry
breaking field, and define a state of the infinite system by
\begin{equation}
 \tilde{\omega}(\cdots)
 := \lim_{B\downarrow0}\lim_\TDL
 (\Pz(B),(\cdots)\,\Pz(B)).
\end{equation}
We expect that the state $\tilde{\omega}(\cdots)$ is
identical to $\omega_{\theta=0}(\cdots)$ 
(and to  $\omega^{\rm ergodic}_{\theta=0}(\cdots)$).
The existence of a symmetry breaking in the state
$\tilde{\omega}(\cdots)$ (under the assumption that there is a
long range order) was proved in 
\cite{KaplanHorschLinden,KomaTasaki}.
\subsection{``Low-lying eigenstates'' in finite systems}
\label{SecFS}
Let us discuss another implication of our theorem,
which have  direct relevance to 
numerical diagonalization approaches to quantum many-body 
systems.

Consider a lattice $\Lambda$ with $N$ sites, and a quantum many-body
system on it which satisfies the assumptions i)-vi) in Section~\ref{SecMain}.
Note that we  require the higher $U(1)\times{\bf Z}_2$ symmetry.
We denote by $\EL^{(0)}$ the ground state energy of the Hamiltonian $\ham$.
Then we can state the following.
\begin{coro}
For each nonvanishing integer $M$ 
which satisfies the bound (\ref{M/N<}), one can find an eigenstate 
$\Phi_\Lambda^{(M)}$ of the Hamiltonian $\ham$.
The state $\Phi_\Lambda^{(M)}$ is orthogonal to the ground state 
$\Pz$, and the states $\Phi_\Lambda^{(M)}$ with 
distinct $M$ 
are orthogonal to each other.
The energy eigenvalue $\EL^{(M)}$ of the state $\PL^{(M)}$ 
satisfies the bound
\begin{equation}
\EL^{(M)}-\EL^{(0)}\le c_3\frac{M^2}{N},
\label{llenergy}
\end{equation}
where $c_3$ is the constant introduced in Theorem~\ref{smallMtheorem}.
\label{llecoro}
\end{coro}
\begin{proof}{Proof}
Since the reference state $\PL$ is an eigenstate of $\gen$, the 
commutation relations (\ref{commutation2}) imply that the 
variational states 
$\PsL^{(M)}$ of (\ref{Psi}) are orthogonal to the reference state 
$\PL$.
Similarly we see that $\PsL^{(M)}$ with distinct $M$ 
are orthogonal to each other.
The desired result is then a consequence of the variational 
principle 
and Theorem~\ref{smallMtheorem}, in which we set $\PL=\Pz$.%
\end{proof}

Consider a sequence of finite lattices $\{\Lambda\}$ which tends
to ${\bf Z}^d$.
Suppose that, for each $\Lambda$, we have a quantum many-body system
which satisfies the assumptions 
i)-vi) with constants $h$, $o$, $r$, $\mu$, and $\gamma$ which are
independent of $\Lambda$.
Then the above Corollary shows that there exist ever increasing numbers of 
``low-lying eigenstates'' whose excitation energies are bounded from
above by a constant times $N^{-1}$.
Such a finite size scaling behavior is characteristic for systems with
a continuous symmetry breaking, and may be regarded as a criterion for 
detecting the existence of a symmetry breaking from numerical diagonalization
in a series of finite systems.
Such an approach has been taken in \cite{Bernu,Azaria,Momoi1}.
We stress that this is the first time that the existence of ever increasing numbers
of ``low-lying eigenstates'' with particular finite size scaling behavior has been
proved.

Nambu-Goldstone excitations associated with the symmetry breaking should
also be observed as ``low-lying excited states'' in finite systems.
According to the common wisdom, the excitation energy of a Nambu-Goldstone 
excitation should be at least of order $L^{-2}$, where $L$ denotes the 
linear dimension of the system.
Therefore the above Corollary guarantees that, in the dimensions
not less than three, the ``low-lying eigenstates''
which are ``parts'' of infinite volume ground states have much lower
energies than Nambu-Goldstone excitations, and can be distinguished from
the latter.

\bigno
{\bf Remark:}
One might wonder if one can get conclusions similar to 
Corollary~\ref{llecoro} from Theorem~\ref{generalMtheorem},
our first theorem on the ``low-lying states''.
By following exactly the same logic as the above,
one finds that Theorem~\ref{generalMtheorem} implies the existence of 
``low-lying eigenstates'' whose excitation energies are (at most) of order 1.
Unfortunately, such information alone is not at all meaningful.
In any system (with or without symmetry braking),  one
can construct excited states with the excitation energy of order 1 by
simply locating  a finite number of ``local defects'' into a (finite volume)
ground state.
We wish to thank Tom Kennedy and the referee for clarifying
this point, which was not properly treated in the earlier version of the
present paper.

In this sense, Theorem~\ref{generalMtheorem} carries no
nontrivial information about
the low-lying spectrum of the finite volume Hamiltonian.
As we have stressed before, however, the true value of 
this theorem is that it establishes that the particular class of states
(like $\PsL^{(M)}$ or $\Xi_\Lambda^{(k)}$) are ``low-lying'' and
converge to infinite-volume ground states.
\Section{Examples}
\label{Sec3}
In this section, we shall discuss some examples to 
which our general results apply.
Although we only discuss selected models representing 
typical situations, the reader can easily extend the following 
analysis to much wider class of quantum many-body problems.
\subsection{Ising model under transverse field}
We shall briefly discuss the Ising model under transverse magnetic 
field with the Hamiltonian (\ref{Ising1}) considered in Section~\ref{SecEx}.
If the field $B$ is smaller than the critical value, the unique ground 
state $\Pz$ is expected to exhibit an ``obscured symmetry breaking'' 
in the sense that the relations $\bktz{\op}=0$ and 
$\bktz{(\op)^2}\ge(m(B)N)^2$ hold, where the order operator $\op$ is 
defined in (\ref{opIsing}).
This can be proved rigorously for sufficiently small $B$.

Then Horsch and von der Linden's theorem 
(Theorem~\ref{M=1theorem}) ensures that there exists a 
``low-lying eigenstate'' whose excitation energy is bounded 
from 
above by a constant times $N^{-1}$.
Note that the theorem does not reproduce the expected exponential 
decay (\ref{expdecay}) of the excitation energy.
\subsection{Heisenberg antiferromagnet 
with N\'eel order}
\label{SecHAF}
We discuss the Heisenberg quantum antiferromagnetic spin system, 
which is a typical model with a spontaneously broken continuous 
symmetry.
Let $\Lambda$ denote the 
$d$-dimensional $L\times\cdots\times L$ hypercubic lattice with 
periodic boundary conditions, where $L$ is an even integer.
With each site $x\in\Lambda$, we associate
the spin operators $(S^{(1)}_x,S^{(2)}_x,S^{(3)}_x)$
for spin $S=1/2,1,3/2,\ldots$.
The Hamiltonian (\ref{ham}) is defined by the local Hamiltonian
\begin{equation}
h_x = \frac{1}{2}\sum_{y;|x-y|=1}
\rbk{S^{(1)}_xS^{(1)}_y+S^{(2)}_xS^{(2)}_y+\lambda 
S^{(3)}_xS^{(3)}_y},
\label{HAF}
\end{equation}
where $0\le\lambda\le1$, and the sum is over 
the sites $y$ neighboring to $x$.
When $L$ is finite, the ground state $\Pz$ of the 
Hamiltonian 
(\ref{HAF}) is rigorously known  \cite{Marshall,LiebMattis,AffleckLieb} 
to be unique and 
satisfies
\begin{equation}
C_\Lambda\Pz=0,
\end{equation}
with $C_\Lambda=\sum_{x\in\Lambda}S^{(3)}_x$.

For $\alpha=1,2$, we define the order operators (\ref{orderOp}) by
the local order operators
\begin{equation}
o^{(\alpha)}_x=\left\{\begin{array}{ll}
S^{(\alpha)}_x&\mbox{if $x\in A$}\\
-S^{(\alpha)}_x&\mbox{if $x\in B$},
\end{array}\right.
\label{orderHAF}
\end{equation}
where we have decomposed $\Lambda$ into two sublattices as 
$\Lambda=A\cup B$ so that for any neighboring sites $x,y$ we have 
either $x\in A$, $y\in B$ or $x\in B$, $y\in A$.

It is expected, and is partially proved by the Dyson-Lieb-Simon 
method and its extensions 
\cite{DysonLiebSimon,Jordao,KennedyLiebShastry1,%
KennedyLiebShastry2,Kubo,KuboKishi,NishimoriKubo,%
NishimoriOzeki,Ozeki} 
that, for any 
$d\geq2$ and $0\le\lambda\le1$, the ground state $\Pz$ 
exhibits a
N\'eel-type long range order.
We expect that the condition (\ref{LRO}) is valid with $\mu>0$
(where $\mu$ depends on $d$ and $\lambda$, but not on the lattice
size $N$).
On the other hand, the absence of  explicit symmetry breaking as in 
(\ref{NOSB}) is obvious from the uniqueness of the ground state.
We thus conclude that there is an ``obscured symmetry breaking''.

Assuming the existence of the N\'eel order (\ref{LRO}), 
we can apply our ``low-lying 
states'' theorems.
The model has a desired $U(1)\times{\bf Z}_2$ symmetry.
We can use 
Theorems~\ref{generalMtheorem} and \ref{smallMtheorem} by noting 
that the present order operators $\op^{(1)}$ and $\op^{(2)}$, along 
with the above defined
$\gen$, satisfy the requirements in 
the theorems with $\gamma=1$.

Then we can make use of the general considerations in Section~\ref{SecFS}, 
and conclude that there are ``low-lying eigenstates'' with
excitation energies not larger than of order $N^{-1}$.
For the $SU(2)$ invariant Heisenberg antiferromagnet with 
$\lambda=1$ in (\ref{HAF}), Momoi \cite{Momoi} constructed some 
additional ``low-lying eigenstates''.

In order to apply the general results in Section~\ref{SecIG}, we need
an extra care.
Since the order operators (\ref{orderHAF}) do not satisfy the 
requirement of the translation invariance, we have to redefine
what we mean by a ``site''.
We group together $2^d$ sites forming a $2\times\cdots\times2$
hypercubic region (i.e., a unit cell), and call such group a ``site''.
After redefining the local Hilbert space, the local Hamiltonian,
and the local order operators according to the new notion of
``sites'', the model satisfies the assumptions of Section~\ref{SecIG}.
We can then form (presumably ergodic) ground states with explicit 
symmetry breaking as 
in (\ref{ergodic}).

We stress that the applicability of our ``low-lying states'' theorems is 
not limited to models on the hypercubic lattice.
For example, the model on the triangular lattice with the same 
Hamiltonian (\ref{HAF}) with $0\le\lambda\le1$, where $y$ is 
summed over nearest neighbor sites of $x$, has been attracting 
considerable 
interest.
(See \cite{Bernu,Azaria,Leung,Momoi1} and many early references therein.)
Anticipating the so-called $120^\circ$ structure, we set the order 
operators as
\begin{equation}
\op^{(1)}=
\sum_{x\in A}
S^{(1)}_x
+\sum_{x\in B}
\rbk{-\frac{1}{2}S^{(1)}_x-\frac{\sqrt{3}}{2}S^{(2)}_x}
+\sum_{x\in C}
\rbk{-\frac{1}{2}S^{(1)}_x+\frac{\sqrt{3}}{2}S^{(2)}_x},
\end{equation}
\begin{equation}
\op^{(2)}=
\sum_{x\in A}
S^{(2)}_x
+\sum_{x\in B}
\rbk{\frac{\sqrt{3}}{2}S^{(1)}_x-\frac{1}{2}S^{(2)}_x}
+\sum_{x\in C}
\rbk{-\frac{\sqrt{3}}{2}S^{(1)}_x-\frac{1}{2}S^{(2)}_x},
\end{equation}
where we have divided the triangular lattice into three sublattices 
$A$, $B$, and $C$, so that neighboring sites $x,y$ always belong to 
different sublattices.
These order operators again satisfy the conditions for the 
theorems with the generator $\gen=\sum_xS^{(3)}_x$.
We can then apply the general discussions in Section~\ref{Sec2}.
\subsection{Bose-Einstein condensation in  hard core Bose gas on  
lattice}
\label{BEsec}
We shall give a brief discussion on the Bose-Einstein condensation problem.
It turns out that, by following the general discussions given in  
Section~\ref{Sec2}, we are naturally led to consider ground states
with unconserved particle number.

Let $\Lambda$ be the $d$-dimensional  $L\times\cdots\times L$
hypercubic 
lattice with periodic boundary conditions, where $L$ is an even 
integer, and 
$d\ge2$.
With each site $x$, we associate the creation operator $a^*_x$ and 
the 
annihilation operator $a_x$ of a spinless boson.
We consider the Hamiltonian (\ref{ham}) defined by
\begin{equation}
h_x = \frac{K}{2}\sum_{y;|x-y|=1}(a^*_xa_y+a^*_ya_x)
+Vn_x(n_x-1),
\label{hamBose}
\end{equation}
where the sum runs over the sites neighboring to $x$, and 
$n_x=a^*_xa_x$ denotes the number operator.

We shall take the limit of infinitely
 large on-site repulsion $V\toinf$ 
before 
the infinite volume limit, and restrict ourselves to the states with 
finite 
energies (in a finite volume).
This defines the so-called hard core Bose gas.

It is well-known that the hard core Bose gas on a lattice is 
equivalent to the 
$S=1/2$ quantum XY model on the same lattice 
\cite{MatsubaraMatsuda}.
Based on the equivalence and an extension of the infrared bound 
method of 
Dyson, Lieb and Simon \cite{DysonLiebSimon}, it was proved by 
Kennedy, Lieb and Shastry \cite{KennedyLiebShastry2}, 
and by 
Kubo and Kishi \cite{KuboKishi} 
that the present model exhibits a Bose-Einstein condensation 
in the 
following sense.
Let $\Pz$ be the unique ground state of the Hamiltonian 
(\ref{hamBose}) with 
the particle number equal to $N=L^d/2$.
Define the order operators (\ref{orderOp}) by
\begin{equation}
o^{(1)}_x=\calP\frac{a^*_x+a_x}{2}\calP,\quad
o^{(2)}_x=\calP\frac{a^*_x-a_x}{2i}\calP,
\label{boseOrder}
\end{equation}
where $\calP$ is the projection operator onto the space of finite energy 
states, i.e., $\Phi$ such that $n_x(1-n_x)\Phi=0$ for any $x$.
Then the result of \cite{KennedyLiebShastry2,KuboKishi} is that the  
condition 
of the long range order (\ref{LRO}) holds with a finite $\mu$
which is independent of the lattice size $N$.
On the other hand the absence of explicit symmetry breaking 
as in (\ref{NOSB}) is 
manifest since the state $\PL$ has a fixed particle number.
We see that there is an ``obscured symmetry breaking''.

It is not hard to see that we can apply our 
Theorems~\ref{generalMtheorem} and \ref{smallMtheorem} to this 
situation.
For this, we replace the Hamiltonian (\ref{hamBose}) with
$\calP h_x\calP$.
Note that the latter is a bounded operator, while the former is
unbounded.
Since we are only dealing with states with finite energies,
this replacement does not change any physics.
By using the replaced Hamiltonian and the order operators
(\ref{boseOrder}), we find that the model has a 
desired $U(1)\times{\bf Z}_2$ symmetry.
The relevant $U(1)$ symmetry is that for the quantum mechanical 
phase 
generated by $\gen=\sum_{x\in\Lambda}(n_x-1/2)$,
and the ${\bf Z}_2$ symmetry is the hole-particle symmetry.

Note that the ``low-lying states'' in the present model have different particle
numbers than the ground state.
The ``low-lying state'' (\ref{Xidef}) with explicit symmetry breaking
can be written as
\begin{equation}
\Xi^{(k)}_\Lambda=\frac{1}{\sqrt{2k+1}}\cbk{
\Pz+\sum_{M=1}^k\rbk{
\frac{\calP(\sum_{x\in\Lambda}a^*_x)^M\Pz}
{\norm{\calP(\sum_{x\in\Lambda}a^*_x)^M\Pz}}
+\frac{(\sum_{x\in\Lambda}a_x)^M\Pz}
{\norm{(\sum_{x\in\Lambda}a_x)^M\Pz}}
}}.
\label{XiforBose}
\end{equation}
Consequently the  (presumably ergodic) ground states 
$\omega_\theta(\cdots)$ have a peculiar feature 
that they are constructed by summing up the states with different 
particle numbers as in (\ref{XiforBose}).
We further find from (\ref{ergodicorder}) that the state has 
nonvanishing 
expectation values of the creation and annihilation operators,
for example, as
\begin{equation}
\omega_{\theta=0}(a^*_x)=\omega_{\theta=0}(a_x)\ge\sqrt{2}o\mu,
\end{equation}
for any $x$.

In a theoretical treatment of the Bose-Einstein condensation, it is  
standard to 
consider states without particle number conservation and 
with nonvanishing 
expectation values for creation and annihilation operators.
Usually such states are introduced within the framework of certain mean-field
theory.
We have seen that such states arise naturally if one tries to 
consider ergodic 
infinite volume ground states.

\bigskip\noindent
{\bf Remark:}
Since it is physically meaningless to compare the energies of two 
states with 
different particle numbers, the existence of ``low-lying states'' in the 
present 
situation has less physical significance.
A more important fact is that the states $\omega_\theta(\cdots)$ 
are really 
infinite volume ground states.
This point requires further discussions.
({\em Note added in August 1997:}
As we have noted in the footnote \ref{BEfn}, states with unconserved 
particle number is physically meaningless.
We now regard this remark as rather technical.)

A physically natural setup in the present problem
is to consider a finite system with a fixed particle number.
Then one can add an extra term $\nu\sum_{x\in\Lambda}n_x$ 
to the Hamiltonian without changing any physics.
If we were to consider states without fixed particle numbers,
Definition~\ref{DEFgs} is clearly not adequate since it is
sensitive to the value of the ``chemical potential''
$\nu$.
Better definitions in the situations without particle number conservation
are those of $\calG_1$ or 
$\calG_2$ in Appendix~\ref{APgs}, with allowed perturbations (local operators 
$A$ in 
the former, and maps $T$ in the latter) restricted to those 
preserve the 
particle number.
This definition is clearly independent of the value of $\nu$.
To prove that the above $\omega_\theta(\cdots)$ is a ground state in this 
sense, it 
suffices to use the relations between different definitions 
(Proposition~\ref{G3}) along with the fact that 
$\omega_\theta(\cdots)$ is a 
ground state (in the sense of Definition~\ref{DEFgs} or $\calG_3$) when 
$\nu=0$.

\subsection{Superconductivity in lattice electron systems}
We shall discuss  applications of our theorems to 
lattice electron problems.
A class of possible applications deals with magnetic ordering in an
electron 
model.
Since such problems can be treated in exactly the same manner as 
the quantum spin systems discussed previously, we leave the 
details to interested readers.
We shall concentrate on a symmetry breaking intrinsic to 
interacting
electron systems, namely, superconductivity.

Consider an electron system on a finite lattice $\Lambda$,
and denote by $ c^*_{x\sigma}$ and $ c_{x\sigma}$ the creation and 
annihilation 
operators, respectively, of an electron at site $x$ with spin 
$\sigma=\uparrow,\downarrow$.
We consider a Hamiltonian which commutes with the total electron
number 
\begin{equation}
N_{\rm e}=\sum_x n_{x\uparrow}+n_{x\downarrow},
\end{equation}
where $n_{x\sigma}= c^*_{x\sigma} c_{x\sigma}$.
A typical example is the so-called Hubbard model.
(See, for example, \cite{Lieb,Montorsi}.)

A class of Hubbard models with attractive interactions is believed 
to exhibit superconductivity in their ground states\footnote{
As a straightforward consequence of the Lieb's theorem \cite{Lieb1}, one 
finds that some attractive Hubbard models exhibit an off-diagonal
long range order \cite{ShenQiu}.
These models, however, do not fit into the present discussion
since the order operator (accidentally) commutes with the
Hamiltonian.
The same comment applies to the solvable models
of \cite{Essler}.
}.
It is also expected that certain Hubbard models with repulsive 
interaction also exhibit superconductivity.
The latter possibility is interesting not only because of its possible 
connection with high-$T_{\rm c}$ superconductivity, but as a new 
type of collective phenomena in strongly interacting electron 
systems.

A standard superconducting phase can be characterized by a 
condensation of certain electron pairs, which manifests itself as an
(off-diagonal)
long range order in electron pairing correlation function.
For example the condensation of singlet pairs can
be measured as a long range order (\ref{LRO}) with respect to the 
order operators defined by
\begin{equation}
o_x^{(1)} = 
\frac{1 }{2}\rbk{p_xc^*_{x\uparrow}c^*_{x\downarrow}-
\overline{p_x}c_{x\uparrow}c_{x\downarrow}}, 
\end{equation}
\begin{equation}
o_x^{(2)} = 
\frac{1}{2i}\rbk{p_xc^*_{x\uparrow}c^*_{x\downarrow}+
\overline{p_x}c_{x\uparrow}c_{x\downarrow}}, 
\end{equation}
where $p_x$ (with $|p_x|=1$) is a certain phase factor. 
It is easily checked that the model has a $U(1)$ symmetry and
satisfies the conditions for 
Theorem~\ref{generalMtheorem} with $\gen =(N_{\rm e}-N)/2$,
where $N$ denotes the number of sites in $\Lambda$.
We can then follow the general discussions in Section~\ref{Sec2}
and construct (presumably) ergodic ground states with 
an explicit $U(1)$ symmetry breaking.
As in the Bose-Einstein condensation problem, the ground states
do not conserve particle numbers.
Except for  half-filled models with special Hamiltonians, the 
models do not have a $U(1)\times{\bf Z}_2$ symmetry necessary to 
apply Theorem~\ref{smallMtheorem}.

We remark that it is possible to treat other types of pairing with 
some extra care.
To treat triplet pairing, for example, we first redefine what we
mean by sites of the lattice.
 We divide the lattice $\Lambda$ into a disjoint union of
non-overlapping pairs of sites.
 We then regard each pair $\{x,y\}$ as a ``site'' of the lattice.
 The local order parameters to measure a possible condensation
of triplet pairs are defined by summing up the following local
order operators over all the ``sites''.
\begin{equation}
o_{\{x,y\}}^{(1)} =
\frac{1}{2}\rbk{c^*_{x\uparrow}c^*_{y\downarrow}
+c^*_{x\downarrow}c^*_{y\uparrow}
-c_{x\uparrow}c_{y\downarrow}
-c_{x\downarrow}c_{y\uparrow}}, 
\end{equation}
\begin{equation}
o_{\{x,y\}}^{(2)} = 
\frac{1}{2i}\rbk{c^*_{x\uparrow}c^*_{y\downarrow}
+c^*_{x\downarrow}c^*_{y\uparrow}
+c_{x\uparrow}c_{y\downarrow}
+c_{x\downarrow}c_{y\uparrow}}
\end{equation}
We again set $\gen =(N_{\rm e}-N)/2$, and apply 
Theorem~\ref{generalMtheorem} to control ``low-lying states''.

\bigskip\noindent
{\bf Remark:}
The lattice fermion problems considered here are different from
other examples in that the corresponding Hilbert space is not a
simple tensor product of local Hilbert spaces as in (\ref{Hilb})
or (\ref{Hilb2}). 
 This difference causes no problem for proving our ``low-lying 
states'' theorems since we only make use of some
commutation relations between operators in the proof.
 But some results about infinite volume states, which are mainly
quoted from literature in Section~\ref{SecIG} and Appendix~\ref{APgs}, 
may not apply.

\subsection{$S=1$ antiferromagnetic chain}
A rather interesting application of the ``low-lying states'' theorem
can be found in the problem related
to the so-called Haldane gap.
Let $\Lambda$ be the one-dimensional open chain $\{1,2,\ldots,N\}$.
With each site $x\in\Lambda$, we associate the three dimensional 
Hilbert space for an $S=1$ quantum spin, and denote by 
$(S^{(1)}_x,S^{(2)}_x,S^{(3)}_x)$ the corresponding spin operators.
We consider the Hamiltonian
\begin{equation}
H_\Lambda = \sum_{x=1}^{N-1}
\rbk{S^{(1)}_xS^{(1)}_{x+1}+S^{(2)}_xS^{(2)}_{x+1}+\lambda 
S^{(3)}_xS^{(3)}_{x+1}}
+D\sum_{x=1}^N (S^{(3)}_x)^2,
\label{S=1Ham}
\end{equation}
where $\lambda$ and $D$ are parameters.

Haldane \cite{Haldane} argued that, in a 
finite range of 
the parameter space including the Heisenberg point $\lambda=1$, 
$D=0$, the model is in an exotic phase (now called 
the ``Haldane phase'') where the unique infinite volume
ground state is accompanied by a finite excitation gap.
This was quite surprising since the Heisenberg antiferromagnetic 
chain with $S=1/2$ is known to have vanishing gap from the Bethe 
ansatz solution.
 (See also \cite{AffleckLieb}.)
Haldane's prediction was that the gapful Haldane phase exists if and 
only if the spin $S$ is an integer.

The existence of the Haldane phase in $S=1$ chains has been 
proved rigorously only in the exactly solvable VBS model  
\cite{AffleckKennedy}, its 
non-$SU(2)$-invariant extensions \cite{Fannes}, and perturbations to 
the dimerized VBS model \cite{KennedyTasaki}.
The ref. \cite{Fannes} contains a general treatment of the VBS-type 
models, and the $S=1$ model mentioned here is one of the examples.
The $S=1$ model of \cite{Fannes} and 
the method of constructing the ground state
(but not the proof of the existence of a gap)
were rediscovered by other authors \cite{Klumper}.

The ground state in the Haldane phase is disordered in the sense that 
the spin-spin correlation functions decay exponentially.
Den~Nijs and Rommelse \cite{denNijs} pointed out that the 
ground state in the Haldane phase of an $S=1$ chain has a ``hidden 
antiferromagnetic 
order''.
For $i=1,2,3$, let the string order operator be
\begin{equation}
\op^{(i)}:=\sum_{x=1}^N S^{(i)}_x\exp[i\pi\sum_{y=1}^{x-1}S^{(i)}_y].
\label{stringoperator}
\end{equation}
If we denote the unique normalized ground state 
for finite $\Lambda$ with $N$ 
sites as $\Pz$, we expect to have
\begin{equation}
\bktz{(\op^{(i)})^2}\ge(\sigma^{(i)} N)^2,
\label{stringorder}
\end{equation}
in the Haldane phase with $\sigma^{(1)}=\sigma^{(2)}>0$ and 
$\sigma^{(3)}>0$.

The condition (\ref{stringorder}) corresponds to the 
antiferromagnetic ordering of spins with $S^{(i)}_x=1$ and 
$S^{(i)}_x=-1$, where spins with $S^{(i)}_x=0$ are inserted randomly 
in 
between them \cite{denNijs,Tasaki}.
Again the existence of the ``hidden antiferromagnetic order'' has 
been established only for special classes of models mentioned above.

Let us consider the following trial states
\begin{equation}
\PsL^{(i)}=\frac{\op^{(i)}\Pz}{\norm{\op^{(i)}\Pz}},
\label{Ktrip}
\end{equation}
for each $i=1,2,3$.
We also introduce the operators
\begin{equation}
U_\Lambda^{(i)}=\exp[i\pi\sum_{x=1}^NS^{(i)}_x],
\end{equation}
which rotates all the spins by $\pi$ around the $i$-th axis.
Note that the unique ground state satisfies 
$U_\Lambda^{(i)}\Pz=\Pz$ for any $i$, and the trial states 
(\ref{Ktrip}) satisfy
\begin{equation}
U_\Lambda^{(i)}\PsL^{(j)}=\left\{
\begin{array}{ll}
\PsL^{(j)}&\mbox{if $i=j$},\\
-\PsL^{(j)}&\mbox{if $i\ne j$}.
\end{array}
\right.
\end{equation}
From the difference of parities, we find that the four states $\Pz, 
\PsL^{(1)}, \PsL^{(2)}$, and $\PsL^{(3)}$ are orthogonal to  each other.

It is not hard to check that we can apply Horsch and von der Linden's 
theorem (Theorem~\ref{M=1theorem})
to this situation.
We find, for each $i=1,2,3$, that the trial states 
$\PsL^{(i)}$ are ``low-lying states''. 
The Hamiltonian (\ref{S=1Ham}) on a finite open chain should 
have (at least) three ``low-lying eigenstates''.
(Again the excitation energies of the ``low-lying eigenstates'' are 
believed 
to decay exponentially in $N$, but the bound in the theorem fails to 
reproduce this.)
These ``low-lying eigenstates'' are nothing but the so-called ``Kennedy 
triplet'' which have been 
observed in exact solutions \cite{AffleckKennedy}, in numerical simulations 
\cite{Kennedy}  and 
in actual experiments in impurity doped samples \cite{Hagiwara}.
The existence of the Kennedy triplet is characteristic in the Haldane 
gap system on a finite open chain.
In a periodic chain, it is believed that the unique  ground state is 
accompanied by a finite excitation gap.

Fundamental connection between the hidden order
(\ref{stringorder}) and the existence
of the low-lying triplet was discussed by Kennedy and Tasaki 
 \cite{KennedyTasaki} from the view point of the 
``hidden ${\bf Z}_2\times{\bf Z}_2$ symmetry breaking''.
Our remark here is that this connection can be made (formally) 
explicit at least in one direction\footnote{
That the existence of a hidden antiferromagnetic 
order should imply the existence of
the low-lying triplet was pointed out to one of the authors (H.T.)
by Ian Affleck in July 1992.
The present work initially emerged from an attempt to
look for a proof of his claim, although the main interest
of the authors has shifted in the long run to  problems with
continuous symmetry breaking.
}.

The present example is different from the others in that the three 
``low-lying states'' and the unique (finite volume) ground state 
converge to a unique infinite volume ground state.
This is related to the non-local nature of the order operators.
See \cite{AffleckKennedy,Kennedy,KennedyTasaki} for more details.
\Section{Proof of first theorem}
\label{Sec4}
In the present section, we prove Theorem~\ref{generalMtheorem} for 
$M>0$.
Throughout the proof, we drop the subscript $\Lambda$ from 
$\op^\pm, \ham, 
E_\Lambda, \PL, \PsL^{(M)}$, etc.
Our goal is to bound the quantity
\begin{eqnarray}
\Delta^{(M)}
&:=&
\frac{1}{N}\cbk{(\Psi^{(M)},H\,\Psi^{(M)})-E}
\ret
&=&
\frac{\gs{\kminus^M H \kplus^M}-\gs{\kminus^M \kplus^M H}}
{N\gs{\kminus^M \kplus^M}}
\ret
&=&
\sum_{x\in\Lambda}
\frac{\gs{\kminus^M[h_x,\kplus^M]}}
{N\gs{\kminus^M \kplus^M}}.
\label{DeltaDef}
\end{eqnarray}

The final expression in (\ref{DeltaDef}) motivates us to decompose 
the 
operator $\Oplus$ as
\begin{equation}
\Oplus = \Qx+\Rx,
\end{equation}
with
\begin{equation}
\Qx:=\sum_{y\not\in\calS_x}o^+_y,\quad
\Rx:=\sum_{y\in\calS_x}o^+_y.
\end{equation}
Note that we have $[\Qx,h_x]=0$ and $[\Qx,\Rx]=0$ from the 
assumptions ii) 
and i), respectively.

Although $O^+$ does not commute with the local Hamiltonian
$h_x$, but $\Qx$ does.
This means that it is easier for us to treat $\Qx$ than $O^+$.
A key observation for the proof is that the difference between $O^+$ and $\Qx$,
which is denoted as $\Rx$, is small compared to $O^+$.
The following useful lemma, for example, makes use of this fact.

\begin{lemma}
Suppose that the conditions (\ref{N>}) and (\ref{M<}) for $N$ and $M$ 
are 
satisfied.
Then for $k=1,2,\ldots,M$, we have
\begin{equation}
\frac{\gs{\KQsx^{M-k}\KQx^{M-k}}}
{\gs{\KQsx^M\KQx^M}}
\le (\mu o N)^{-2k}.
\label{QQBound}
\end{equation}
\label{QQLemma}
\end{lemma}
We shall prove the lemma at the end of the present section.

By using the expansion formula
\begin{equation}
\kplus^M=\sum_{k=0}^M\Mk\KQx^{M-k}\KRx^k,
\label{OExp}
\end{equation}
we get
\begin{eqnarray}
&&\gs{\kminus^M[h_x,\kplus^M]}
\ret
&=&
\sum_{k=0}^M\sum_{\ell=0}^M\Mk\Ml
\gs{\KQsx^{M-k}\KRsx^k[h_x,\KQx^{M-\ell}\KRx^\ell]}
\ret
&=&
\sum_{k=0}^M\sum_{\ell=1}^M\Mk\Ml
\gs{\KQsx^{M-k}\KRsx^k[h_x,\KRx^\ell]\KQx^{M-\ell}}.
\label{OhO1}
\end{eqnarray}

The Schwartz inequality and the definition (\ref{DEFnorm}) of the 
operator 
norm yield the following useful bounds
\begin{eqnarray}
\abs{\gs{A^*BC}}
&\le&
\sqrt{\gs{A^*A}\gs{C^*B^*BC}}
\ret
&\le&
\norm{B}\sqrt{\gs{A^*A}\gs{C^*C}},
\label{ABC}
\end{eqnarray}
for general operators $A$, $B$, and $C$.

By applying (\ref{ABC}) to (\ref{OhO1}) and noting that ii) and iii) 
imply
$\norm{\KRsx^k[h_x,\KRx^\ell]}\le 2h(2ro)^{k+\ell}$,
we get
\begin{eqnarray}
&&\abs{\frac{\gs{\kminus^M[h_x,\kplus^M]}}{\gs{\KQsx^M\KQx^M}}}
\ret
&\le&
\sum_{k=0}^M\sum_{\ell=1}^M\Mk\Ml2h(2ro)^{k+\ell}
\frac{\sqrt{\gs{\KQsx^{M-k}\KQx^{M-k}}\gs{\KQsx^{M-\ell}\KQx^{M-
\ell}}}}
{\gs{\KQsx^{M}\KQx^{M}}}
\ret
&\le&
2h\sum_{k=0}^M\sum_{\ell=1}^M\Mk\Ml(2ro)^{k+\ell}(\mu o N)^{-
(k+\ell)}
\ret
&=&
2h\rbk{1+\frac{2r}{\mu N}}^M\cbk{\rbk{1+\frac{2r}{\mu N}}^M-1}
\ret
&\le&
2h \exp\sbk{\frac{2r}{\mu}\frac{M}{N}}
\cbk{\exp\sbk{\frac{2r}{\mu}\frac{M}{N}}-1}
\ret
&\le&
2h e^{\mu/4}\frac{8r(e^{\mu/4}-1)}{\mu^2}\frac{M}{N}
\ret
&=&
16rh\mu^{-2}(e^{\mu/2}-e^{\mu/4})\frac{M}{N},
\label{OhO2}
\end{eqnarray}
where we have used the bounds (\ref{QQBound}) and (\ref{M<}).

Again using (\ref{OExp}), we get
\begin{eqnarray}
&&\gs{\kminus^M\kplus^M}
\ret&=&
\gs{\KQsx^M\KQx^M}
\ret&+&
{\sum_{k,\ell}}'\Mk\Ml\gs{\KQsx^{M-k}\KRsx^k\KRx^\ell\KQx^{M-
\ell}},
\end{eqnarray}
where the summation in the right-hand side runs over all 
$k,\ell=0,1,\ldots,M$ except for $k=\ell=0$.
From (\ref{ABC}) and (\ref{QQBound}), we get
\begin{eqnarray}
&&\abs{\frac{\gs{\kminus^M\kplus^M}}{\gs{\KQsx^M\KQx^M}}}
\ret
&\ge&
1-{\sum_{k,\ell}}'\Mk\Ml(2ro)^{k+\ell}
\frac{\sqrt{\gs{\KQsx^{M-k}\KQx^{M-k}}\gs{\KQsx^{M-\ell}\KQx^{M-
\ell}}}}
{\gs{\KQsx^M\KQx^M}}
\ret
&\ge&
1-\cbk{\rbk{1+\frac{2r}{\mu N}}^{2M}-1}
\ret
&\ge&
2-e^{\mu/2}.
\label{OOQQ}
\end{eqnarray}
Note that, since $0\le\mu\le1$, we have $2-e^{\mu/2}\ge2-
\sqrt{e}>0$.

By combining (\ref{DeltaDef}), (\ref{OhO2}), and (\ref{OOQQ}), we 
finally get
\begin{equation}
\abs{\Delta^{(M)}}
\le
16rh\frac{e^{\mu/2}-e^{\mu/4}}{\mu^2(2-e^{\mu/2})}\frac{M}{N}
=c_1\frac{M}{N}.
\end{equation}

\begin{proof}{Proof of Lemma~\ref{QQLemma}}
We write $a_m:=\gs{\KQsx^m\KQx^m}$.
We will prove that for $m=1,2,\ldots,M$, we have
\begin{equation}
\frac{a_m}{a_{m-1}}\ge(\mu o N)^2.
\label{a/a}
\end{equation}
Then the desired bound (\ref{QQBound}) follows by multiplying 
(\ref{a/a}) 
with $m=M-k+1,M-k+2,\ldots,M$.

We start by evaluating $a_1$ as
\begin{eqnarray}
a_1 &=& \gs{(\Ominus-\Rsx)(\Oplus-\Rx)}
\ret
&\ge& \gs{\Ominus\Oplus}-2(2o)^2rN
\ret
&=& \frac{1}{2}\cbk{\gs{\Ominus\Oplus}+\gs{\Oplus\Ominus}
+\gs{[\Ominus,\Oplus]}}-8o^2rN
\ret
&\ge& \gs{\kone^2}+\gs{(O^{(2)})^2}-2o^2(1+4r^2)N,
\end{eqnarray}
where we have used i), ii), iii) to bound the norm of the 
commutators.

By substituting the assumption (\ref{LRO}) on the existence of a long 
range 
order, and the bound (\ref{N>}) for $N$, we get
\begin{eqnarray}
a_1&\ge&2(\mu o N)^2 \cbk{1-\frac{1+4r^2}{\mu^2N}}
\ret
&\ge&2(\mu o N)^2 \cbk{1-\frac{1+4r^2}{16r^2}}.
\label{a1Bound}
\end{eqnarray}
Since $r\ge2$, we have shown that $a_1>0$.

Next we use the Schwartz inequality to get
\begin{eqnarray}
(a_{m-1})^2 &=& \gs{\KQsx^{m-2}\Qsx\KQx^{m-1}}^2
\ret
&\le& \gs{\KQsx^{m-2}\KQx^{m-2}}\gs{\KQsx^{m-1}\Qx\Qsx\KQx^{m-
1}}
\ret
&=&
a_{m-2}\cbk{\gs{\KQsx^m\KQx^m}+\gs{\KQsx^{m-
1}[\Qx,\Qsx]\KQx^{m-1}}}
\ret
&\le&
a_{m-2}\cbk{a_m+4o^2Na_{m-1}},
\label{am-1}
\end{eqnarray}
where the final inequality follows from (\ref{ABC}).

Assuming that $a_{m-2}\ne0$ and $a_{m-1}\ne0$ (which is true for 
$m=2$), 
we find from (\ref{am-1}) that
\begin{equation}
\frac{a_m}{a_{m-1}}\ge\frac{a_{m-1}}{a_{m-2}}-4o^2N.
\label{recBound}
\end{equation}

The rest of the proof is easy.
Assume that, for a fixed $m$, $a_{m'}\ne0$ for all $m'<m\le M$.
Then by summing up (\ref{recBound}) and using the bounds 
(\ref{a1Bound}), 
(\ref{M<}) and $r\le2$, we see that
\begin{eqnarray}
\frac{a_m}{a_{m-1}}&\ge&a_1-4o^2N(m-2)
\ret
&\ge&2(\mu o N)^2\cbk{1-\frac{1+4r^2}{16r^2}-\frac{2(m-
2)}{\mu^2N}}
\ret
&\ge&2(\mu o N)^2\cbk{1-\frac{1+4r^2}{16r^2}-\frac{1}{4r}}
\ret
&\ge&(\mu o N)^2,
\end{eqnarray}
and hence $a_m\ne0$.
By proceeding inductively, we see that the desired bound (\ref{a/a}) 
holds for 
$m=1,2,\ldots,M$.%
\end{proof}
\Section{Proof of second theorem}
\label{Sec5}
In the present section, we shall prove 
Theorem~\ref{smallMtheorem}.
Again we fix $\Lambda$ and drop the subscript $\Lambda$.

Let us define
\begin{equation}
b_m:=\frac{2^{2m}(m!)^2}{(2m)!}\gs{\kone^{2m}},
\label{bmDef}
\end{equation}
which satisfies the following useful inequalities.

\begin{lemma}
We have
\begin{equation}
\frac{1}{2(oN)^2}\le\frac{b_{m-1}}{b_m}\le\frac{1}{(\mu oN)^2}.
\label{b/b}
\end{equation}
\label{easyLemma}
\end{lemma}
\begin{proof}{Proof}
By using the Schwartz inequality we get
\begin{equation}
\gs{\kone^{2m-2}}^2=\gs{\kone^m\kone^{m-2}}^2
\le\gs{\kone^{2m}}\gs{\kone^{2m-4}},
\label{OO1}
\end{equation}
which, with $\gs{\kone^2}>0$, proves inductively 
$\gs{\kone^{2m}}>0$ for any 
$m$.
By rearranging (\ref{OO1}), we get
\begin{equation}
\frac{\gs{\kone^{2(m-1)}}}{\gs{\kone^{2m}}}
\le \frac{\gs{\kone^{2(m-2)}}}{\gs{\kone^{2(m-1)}}}
\le\cdots\le
\frac{1}{\gs{\kone^2}}\le\frac{1}{(\mu oN)^2},
\end{equation}
where we used (\ref{LRO}).
We also note that the definition (\ref{DEFnorm}) implies
\begin{equation}
\gs{\kone^{2m}}\le\norm{\oone}^2\gs{\kone^{2(m-1)}}
\le(oN)^2\gs{\kone^{2(m-1)}}.
\label{OO2}
\end{equation}
By substituting (\ref{OO1}) and (\ref{OO2}) into
\begin{equation}
\frac{b_{m-1}}{b_m}=
\frac{2m(2m-1)}{(2m)^2}\frac{\gs{\kone^{2(m-
1)}}}{\gs{\kone^{2m}}},
\end{equation}
which follows from the definition (\ref{bmDef}), we get 
(\ref{b/b}).%
\end{proof}

In the present proof, we shall bound various quantities
in terms of $b_M$.
One of the main ingredients in the proof is the following 
Lemma~\ref{hardLemma}, which allows us to approximate 
expectation 
values including $O^\pm$ with those including the self-adjoint 
operator 
$\oone$.
See equation (\ref{useLemma}) for a typical situation to which we
apply the lemma.

Let $A$ be an operator written as $A=\sum_{x\in\Lambda}a_x$, 
where $[a_x, o_y^{(\alpha)}]=0$ for 
 $\alpha=1,2$ if $y\not\in\calS_x$, and
 $\norm{a_x}\le a$ for any $x$ with an $x$-independent 
finite constant $a$.
The support set $\calS_x$ is the same for that of $h_x$.
In the following we set $\osig{}=\Oplus$ or $\Ominus$ depending on 
$\sigma=+1$ or $-1$.

\begin{lemma}
Let $K,L$ be nonnegative integers which satisfy
\begin{equation}
\frac{48r}{\mu^2}\frac{K+L}{N}
+\frac{3\gamma}{2o^2\mu^2}\frac{(K+L)^3}{N^2} \le 1.
\label{K+Lcond}
\end{equation}
We assume that $A$ satisfies
\begin{equation}
[C,\kplus^{K-L}A]=0,
\label{Acond}
\end{equation}
where $C$ is the generator of the $U(1)$ symmetry.
(For $K-L<0$, we set $\kplus^{K-L}=\kminus^{L-K}$.)
The relation (\ref{Acond}) essentially means that $A$ consists of 
$(K-L)$ 
lowering operators.
Let $\{\sigma_i\}_{i=1,\ldots,K+L}$ be such that $\sigma_i=\pm1$ 
and 
$\sum_{i=1}^{K+L}\sigma_i=K-L$.
Then for any integer $k$ with $0\le k\le K+L$, we have
\begin{eqnarray}
&&\abs{
\gs{\rbk{\prod_{i=1}^k\osig{i}}A\rbk{\prod_{i=k+1}^{K+L}\osig{i}}}
-\frac{2^{2J}(J!)^2}{(2J)!}\gs{\kone^J\cbk{A\kplus^{(K-L)}}\kone^J}
}
\ret
&&\le\delta(A;K,L),
\label{dBound1}
\end{eqnarray}
where $J=\min\{K,L\}$, and
\begin{equation}
\abs{
\gs{\rbk{\prod_{i=1}^k\osig{i}}A\rbk{\prod_{i=k+1}^{K+L}\osig{i}}}
}
\le 3\delta(A;K,L).
\label{dBound2}
\end{equation}
Here $\delta(A;K,L)$ is given by
\begin{equation}
\delta(A;K,L):=\frac{1}{2}(aN)(2oN)^{|K-L|}b_J,
\label{delta1}
\end{equation}
for general $A$.
For $A=1$, in which case only $K=L$ is allowed, we can set
\begin{equation}
\delta(1;K,K)=\frac{1}{2}b_K.
\label{delta2}
\end{equation}
\label{hardLemma}
\end{lemma}
The lemma will be proved after completing the proof of the main 
theorem.

We again want to control the quantity $\Delta^{(M)}$ in 
(\ref{DeltaDef}).
By  using the relation $U\osig{}U^{-1}=-O^{-\sigma}$ (which 
follows form 
(\ref{commutation3})), we find that $\Delta^{(M)}$ can be written in 
terms of 
a double commutator as follows.
(This is reminiscent of the similar representation (\ref{dc})
used in the proof of the simplest ``low-lying states'' theorem
of Section~\ref{SecHV}.)
\begin{eqnarray}
&&
2N\norm{\Psi^{(M)}}^2\Delta^{(M)}
\ret
&=& 2\gs{\kminus^MH\kplus^M}-2E\gs{\kminus^M\kplus^M}
\ret
&=& \gs{\kminus^MH\kplus^M}+\gs{\kplus^MH\kminus^M}
\ret
&& -\gs{\kminus^M\kplus^MH}-\gs{H\kplus^M\kminus^M}
\ret
&=& \gs{[\kminus^M,[H,\kplus^M]]}
\ret
&=& \sum_{m=0}^{M-1} \gs{[\kminus^M,\kplus^m[H,\Oplus]
\kplus^{M-m-1}]}
\ret
&=&
\sum_{m=1}^{M-1}\sum_{\ell=0}^{M-1}\sum_{n=0}^{m-1}
\ret
&& \gs{\kminus^\ell\kplus^n[\Ominus,\Oplus]\kplus^{m-n-
1}[H,\Oplus]
\kplus^{M-m-1}\kminus^{M-\ell-1}}
\ret
&+& \sum_{m=0}^{M-1}\sum_{\ell=0}^{M-1}
\ret
&&\gs{\kminus^\ell\kplus^m[\Ominus,[H,\Oplus]]
\kplus^{M-m-1}\kminus^{M-\ell-1}}
\ret
&+&
\sum_{m=0}^{M-2}\sum_{\ell=0}^{M-1}\sum_{n=0}^{M-m-2}
\ret
&&
\gs{\kminus^\ell\kplus^m[H,\Oplus]\kplus^n[\Ominus,\Oplus]
\kplus^{M-m-n-2}\kminus^{M-\ell-1}}.
\end{eqnarray}
Note that the symmetry vi), along with the relation $UCU^{-1}=-C$ 
(which 
follows from (\ref{commutation1}) and (\ref{commutation3})), 
implies that $C\Phi=0$.
By also using the relation $[\Ominus,\Oplus]=-2\gamma C$, 
and the 
fact that $O^\pm$ are the raising and the lowering operators, 
we can 
bound the above quantity as
\begin{eqnarray}
&&\abs{2N\norm{\Psi^{(M)}}^2\Delta^{(M)}}
\ret
&\le& \sum_{m=1}^{M-1}\sum_{\ell=0}^{M-1}\sum_{n=0}^{m-1}
\ret
&& 2\gamma M
\abs{\gs{\kminus^\ell\kplus^{m-1}[H,\Oplus]
\kplus^{M-m-1}\kminus^{M-\ell-1}}}
\ret
&+& \sum_{m=0}^{M-1}\sum_{\ell=0}^{M-1}
\ret
&&\abs{\gs{\kminus^\ell\kplus^m[\Ominus,[H,\Oplus]]
\kplus^{M-m-1}\kminus^{M-\ell-1}}}
\ret
&+&
\sum_{m=0}^{M-2}\sum_{\ell=0}^{M-1}\sum_{n=0}^{M-m-2}
\ret
&&
2\gamma M
\abs{\gs{\kminus^\ell\kplus^m[H,\Oplus]
\kminus^{M-m-2}\kminus^{M-\ell-1}}}
\ret
&\le&
6\gamma M^4\delta([H,\Oplus];M-2,M-1)
+3M^2\delta([\Ominus,[H,\Oplus]];M-1,M-1),
\label{useLemma}
\end{eqnarray}
where we have used (\ref{dBound2}).
The use of Lemma~\ref{hardLemma} is justified since $M$ satisfies 
(\ref{M/N<}) which (with $K+L\le 2M$) guarantees the condition 
(\ref{K+Lcond}).

By using (\ref{delta1}) in Lemma~\ref{hardLemma}, we get
\begin{eqnarray}
&&\abs{2N\norm{\Psi^{(M)}}^2\Delta^{(M)}}
\ret
&\le& 3\gamma M^4(4rohN)(2oN)b_{M-2}
+\frac{3}{2}M^2(16r^2o^2hN)b_{M-1}
\ret
&\le&\cbk{\frac{24\gamma rh}{o^2\mu^4}\frac{M^4}{N^2}
+\frac{24r^2h}{\mu^2}\frac{M^2}{N}}b_M,
\label{Bunshi}
\end{eqnarray}
where we have used the bound (\ref{b/b}) in Lemma~\ref{easyLemma}
to relate $b_M$ with different $M$.

On the other hand, from (\ref{dBound1}) and (\ref{delta2}), we find
\begin{equation}
\norm{\Psi^{(M)}}^2=\gs{\kminus^M\kplus^M}\ge\frac{1}{2}b_M.
\label{Bunbo}
\end{equation}

By combining (\ref{Bunshi}) and (\ref{Bunbo}), and substituting the 
assumed 
bound (\ref{M/N<}), we finally get
\begin{eqnarray}
\abs{\Delta^{(M)}}
&\le& 
\frac{24r^2h}{\mu^2}\cbk{1+\frac{\gamma}{o^2\mu^2r}\frac{M^2}{N}}
\rbk{\frac{M}{N}}^2
\ret
&\le& c_3\rbk{\frac{M}{N}}^2,
\end{eqnarray}
with $c_3=24r^2h\mu^{-2}\cbk{1+\gamma o^{-2}\mu^{-2}r^{-1}c_2}$.

It remains to prove Lemma~\ref{hardLemma}.
We prepare the following.

\begin{lemma}
Let $K,L$ be nonnegative integers which satisfy (\ref{K+Lcond}).
We assume that the operator $A$ satisfies the conditions of 
Lemma~\ref{hardLemma}.
Let $\{\sigma_i\}_{i=1,\ldots,K+L}$, $\{\tau\}_{i=1,\ldots,K+L}$ 
be such that 
$\sigma_i=\pm1$, $\tau_i=\pm1$, and 
$\sum_{i=1}^{K+L}\sigma_i=\sum_{i=1}^{K+L}\tau_i=K-L$.
Then for any integers $k,\ell$ with $0\le k,\ell\le K+L$, we have
\begin{equation}
\abs{
\gs{\rbk{\prod_{i=1}^k\osig{i}}A\rbk{\prod_{i=k+1}^{K+L}\osig{i}}}
-
\gs{\rbk{\prod_{i=1}^\ell 
O^{\tau_i}}A\rbk{\prod_{i=\ell+1}^{K+L}O^{\tau_i}}}
}
\le
\delta(A;K,L),
\label{dBound3}
\end{equation}
with the same $\delta(A;K,L)$ as in (\ref{delta1}) and (\ref{delta2}).
\label{subLemma}
\end{lemma}

\begin{proof}{Proof of Lemma~\ref{hardLemma} given 
Lemma~\ref{subLemma}}
Let $B$ be an arbitrary operator which satisfies $[B,C]=0$.
Then
\begin{eqnarray}
\gs{\kone^JB\kone^J}
&=& 
\gs{\rbk{\frac{\Oplus+\Ominus}{2}}^JB\rbk{\frac{\Oplus+\Ominus}{2
}}^J}
\ret
&=& \sum_{\tau_1=\pm1}\cdots\sum_{\tau_{2J}=\pm1}
2^{-
2J}\gs{\rbk{\prod_{i=1}^JO^{\tau_i}}B\rbk{\prod_{i=J+1}^{2J}O^{\tau
_i}}}
\ret
&=&
\sum_{\{\tau_i\};\sum\tau_i=0}
2^{-
2J}\gs{\rbk{\prod_{i=1}^JO^{\tau_i}}B\rbk{\prod_{i=J+1}^{2J}O^{\tau
_i}}},
\label{OBO1}
\end{eqnarray}
where the final sum is over all $\tau_i=\pm1$ with 
$\sum_{i=1}^{2J}\tau_i=0$.
The constraint comes from the fact that $\Phi$ is an eigenstate of 
the $U(1)$ 
generator $C$.
Since the number of distinct combinations $\{\tau_i\}$ which 
satisfy the 
constraint is equal to $\JJ=(2J)!/(J!)^2$, we can rewrite 
(\ref{OBO1}) as
\begin{equation}
\JJ^{-1}\sum_{\{\tau_i\};\sum\tau_i=0}
\gs{\rbk{\prod_{i=1}^JO^{\tau_i}}B\rbk{\prod_{i=J+1}^{2J}O^{\tau_i}}
}
=
\frac{2^{2J}(J!)^2}{(2J)!}\gs{\kone^JB\kone^J}
\label{OBO2}
\end{equation}

To prove (\ref{dBound1}), we set $B=A\kplus^{K-L}$, and substitute 
(\ref{OBO2}) into the left-hand side of (\ref{dBound1}) to get
\begin{eqnarray}
&&
\abs{
\gs{\rbk{\prod_{i=1}^k\osig{i}}A\rbk{\prod_{i=k+1}^{K+L}\osig{i}}}
-\frac{2^{2J}(J!)^2}{(2J)!}\gs{\kone^JA\kplus^{K-L}\kone^J}
}
\ret &\le&
\JJ^{-1}\sum_{\{\tau_i\};\sum\tau_i=0}
\ret &&
\abs{
\gs{\rbk{\prod_{i=1}^k\osig{i}}A\rbk{\prod_{i=k+1}^{K+L}\osig{i}}}
- \gs{\rbk{\prod_{i=1}^JO^{\tau_i}}A\kplus^{K-L}
\rbk{\prod_{i=J+1}^{2J}O^{\tau_i}}}
}
\ret &\le&
 \JJ^{-1}\sum_{\{\tau_i\};\sum\tau_i=0}
\delta(A;K,L)
\ret
&=&\delta(A;K,L),
\end{eqnarray}
where we have used (\ref{dBound3}).

To prove (\ref{dBound2}), we simply substitute the estimate
\begin{eqnarray}
&&\abs{
\frac{2^{2J}(J!)^2}{(2J)!}
\gs{\kone^JA\kplus^{K-L}\kone^J}
}
\ret
&& \le \norm{A\kplus^{K-L}}\frac{2^{2J}(J!)^2}{(2J)!}
\gs{\kone^{2J}}
\le
2\delta(A;K,L),
\end{eqnarray}
which follows from the definition of norm (\ref{DEFnorm}), into the bound
(\ref{dBound1}).%
\end{proof}

\begin{proof}{Proof of Lemma~\ref{subLemma}}
The desired bound (\ref{dBound3}) is trivial if $K=L=0$.
We shall prove the bound inductively in $K+L$.

Fix $K,L$ with $K\ge L$ (the bounds for $K<L$ follow from the 
symmetry), and 
assume (\ref{dBound3}) for any nonnegative integers 
$K',L'$ with $K'+L'< K+L$.
Note that we are also allowed to use the resulting bounds 
(\ref{dBound1}) and 
(\ref{dBound2}) for the same $K',L'$.

We first assume $A\ne1$.
We denote by $D$ the left-hand side of (\ref{dBound3}), and bound it 
as
\begin{eqnarray}
D&\le&
\abs{
\gs{\rbk{\prod_{i=1}^k\osig{i}}A\rbk{\prod_{i=k+1}^{K+L}\osig{i}}}
-
\gs{A\rbk{\prod_{i=1}^{K+L}\osig{i}}}
}
\ret&& 
+ \abs{
\gs{A\rbk{\prod_{i=1}^{K+L}\osig{i}}}
-
\gs{A\rbk{\prod_{i=1}^{K+L}O^{\tau_i}}}
}
\ret&&
+ \abs{
\gs{A\rbk{\prod_{i=1}^{K+L}O^{\tau_i}}}
-
\gs{\rbk{\prod_{i=1}^\ell O^{\tau_i}}A
\rbk{\prod_{i=\ell+1}^{K+L}O^{\tau_i}}}
}
\ret
&=:& D_1+D_2+D_3.
\label{Ddecomp}
\end{eqnarray}

The quantity $D_1$ in (\ref{Ddecomp}) can be bounded as
\begin{equation}
D_1 \le
\sum_{j=1}^k
\abs{\gs{\rbk{\prod_{i=1}^{j-1}\osig{i}}[\osig{j},A]
\rbk{\prod_{i=j+1}^{K+L}\osig{i}}}}.
\label{D1<}
\end{equation}
Each term in the sum of (\ref{D1<}) can be bounded by using 
(\ref{dBound2}), 
where we identify $[O^\pm,A]$ as the operator $A$ in 
(\ref{dBound2}).
The condition (\ref{Acond}) is clearly satisfied if we properly
replace $K-L$ by $K-L\mp1$.
To apply (\ref{dBound2}), we rewrite the operator as 
$[O^\pm,A]=\sum_{x\in\Lambda}[\sum_{y\in\calS_x}o^\pm_y,a_x]$,
and note that 
$\norm{[\sum_{y\in\calS_x}o^\pm_y,a_x]}\le4rao$.
When $\sigma_j=+1$, we get
\begin{eqnarray}
&&
\abs{\gs{\rbk{\prod_{i=1}^{j-1}\osig{i}}[\Oplus,A]
\rbk{\prod_{i=j+1}^{K+L}\osig{i}}}}
\ret &\le&
3 \delta([\Oplus,A];K-1,L)
\ret &\le&
\left\{\begin{array}{ll}
\frac{3}{2}(4raoN)(2oN)^{K-L-1}b_L
&\mbox{if $K>L$}\\
\frac{3}{2}(4raoN)(2oN)b_{L-1}
&\mbox{if $K=L$}
\end{array}\right.
\ret &\le&
\frac{3}{2}(4raoN)(2oN)^{K-L+1}b_{L-1},
\end{eqnarray}
for $K\ge L$, where we used the bound (\ref{b/b}) in the case $K>L$.

When $\sigma_j=-1$, we also get
\begin{eqnarray}
&&
\abs{\gs{\rbk{\prod_{i=1}^{j-1}\osig{i}}[\Ominus,A]
\rbk{\prod_{i=j+1}^{K+L}\osig{i}}}}
\ret &\le&
3 \delta([\Ominus,A];K,L-1)
\ret &\le&
\frac{3}{2}(4raoN)(2oN)^{K-L+1}b_{L-1}.
\end{eqnarray}
Since there are at most $(K+L)$ terms in the sum in (\ref{D1<}), we 
find that
\begin{equation}
D_1\le(K+L)\frac{3}{2}(4raoN)(2oN)^{K-L+1}b_{L-1}.
\label{D1<2}
\end{equation}
It is obvious that the quantity $D_3$ satisfies the same bound 
as (\ref{D1<2}).

To evaluate the quantity $D_2$ in (\ref{Ddecomp}), we transform 
$\{\sigma_i\}_{i=1,\ldots,K+L}$ into $\{\tau_i\}_{i=1,\ldots,K+L}$ 
by 
successively exchanging neighboring indices.
We then get
\begin{equation}
D_2\le\sum_{\{\kappa_i\}}
\abs{\gs{A\rbk{\prod_{i=1}^{j-1}O^{\kappa_i}}
[O^{\kappa_j},O^{\kappa_{j+1}}]\rbk{\prod_{i=j+2}^{K+L}O^{\kappa_i}}
}},
\end{equation}
where $\{\kappa_i\}_{i=1,\ldots,K+L}$ is summed over the 
sequence of 
configurations which interpolates between 
$\{\sigma_i\}_{i=1,\ldots,K+L}$ 
and $\{\tau_i\}_{i=1,\ldots,K+L}$, and $j$ (which depends on 
$\{\kappa_i\}_{i=1,\ldots,K+L}$) indicates where the indices are 
exchanged.

By using the commutation relation (\ref{commutation3}), and 
$C\Phi=0$, we 
can further bound $D_2$ as
\begin{eqnarray}
D_2 &\le&
\sum_{\{\kappa_i\}}
2\gamma(K+L)\abs{\gs{A\rbk{\prod_{i=1}^{j-1}O^{\kappa_i}}
\rbk{\prod_{i=j+2}^{K+L}O^{\kappa_i}}}}
\ret
&\le&
\sum_{\{\kappa_i\}}
6\gamma(K+L)\delta(A;K-1,L-1)
\ret
&\le&
3\gamma KL(K+L)(aN)(2oN)^{K-L}b_{L-1},
\label{D2<}
\end{eqnarray}
where we have used the fact that at most $KL$ exchanges are 
necessary to get 
$\{\tau\}_{i=1,\ldots,K+L}$ from $\{\sigma_i\}_{i=1,\ldots,K+L}$, 
and the 
bound (\ref{dBound2}).

By substituting the bounds (\ref{D1<2}) and (\ref{D2<}) into the 
decomposition 
(\ref{Ddecomp}), and using the bounds (\ref{b/b}) and (\ref{delta1}), 
we 
finally get
\begin{eqnarray}
D &\le&
\cbk{24(K+L)ro^2N+3\gamma KL(K+L)}(aN)(2oN)^{K-L}b_{L-1}
\ret
&\le&
\cbk{\frac{48r}{\mu^2}\frac{K+L}{N}
+\frac{6\gamma}{o^2\mu^2}\frac{KL(K+L)}{N^2}}
\delta(A;K,L)
\ret
&\le& \delta(A;K,L),
\end{eqnarray}
where we used the assumption (\ref{K+Lcond}) and $KL\le(K+L)^2/4$.
This proves the desired (\ref{dBound3}) for $A\ne1$.

The case $A=1$ is much easier.
One notes that only $D_2$ is nonvanishing in the decomposition 
(\ref{Ddecomp}).
The similar estimate as the above proves the desired result.%
\end{proof}
\appendix
\Section{Ground states of infinite systems}
\label{APgs}
In the present Appendix, we give mathematically precise definitions of 
ground states in an infinite system, and discuss relations between 
different definitions.
The contents of the present Appendix might be well-known to 
experts, but they have not been written down 
explicitly as far as we know.
We think it would be convenient for the readers to have 
them included in the present paper.

We start by briefly reviewing basic setups in operator algebraic 
approach to quantum systems with infinitely many degrees of 
freedom \cite{BratteliRobinson,Ruelle,Simon}.
For simplicity we shall consider a quantum many-body system 
defined on the $d$-dimensional hypercubic lattice $\Zd$.
With each site $x\in\Zd$, we associate a finite dimensional Hilbert 
space $\calH_x$ which is assumed to be identical to $\calH_o$ 
where 
$o$ is a fixed site (the origin) of $\Zd$.
The Hilbert space corresponding to a finite subset 
$\Omega\subset\Zd$ is
\begin{equation}
\calH_\Omega:=\bigotimes_{x\in\Omega}\calH_x.
\label{Hilb2}
\end{equation}
Let $\calA_\Omega$ denote the set of all the operators on 
$\calH_\Omega$.
The basic object in the operator algebraic approach is the algebra of 
quasi local operators defined as
\begin{equation}
\calA:=\overline{\bigcup_{\Omega}\calA_\Omega},
\label{DEFalgebra}
\end{equation}
where the union is over all the finite subsets $\Omega\subset\Zd$, 
and 
the completion is taken with respect to the norm (\ref{DEFnorm}) 
for local operators.
Note that we have made $\calA$ into a Banach space.

A state $\rho(\cdots)$ is a linear map from  $\calA$ to ${\bf C}$
which 
satisfies $\rho(1)=1$, and $\rho(A^*A)\ge0$ 
for any $A\in\calA$.
It can be shown \cite{BratteliRobinson} that it automatically holds
that $\abs{\rho(A)}\le\norm{A}$, and $\rho(A^*)=\rho(A)^*$.
We denote by $\calE$ the set of all states on $\calA$.
Since $\calE$ is the intersection of the unit sphere of the 
dual space $\calA^*$ and the cone of positive functionals,
Banach-Alaoglu theorem \cite{ReedSimon} implies that $\calE$ is 
compact in the weak-$*$ topology.

The compactness provides us with a useful way of constructing 
states on $\calA$.
Let $\{\Omega_i\}_{i=1,2,\ldots}$ be an arbitrary sequence of finite 
subsets of $\Zd$ which tends to $\Zd$ in the sense of van Hove
\cite{VanHove,Ruelle} as 
$i\toinf$.
For each $i$ we take a state (density matrix) $\rho_i(\cdots)$ on 
the algebra $\calA_{\Omega_i}$.
Since $\rho_i(\cdots)$ can be naturally regarded\footnote{
For $A\in\calA_{\Omega_i}$ and 
$B\in\calA_{{\Omega_i}^c}$ (where
${\Omega_i}^c={\bf Z}^d\backslash\Omega_i$), we set
$\tilde{\rho}_i(AB)=\rho_i(A)\sigma_i(B)$,
where $\sigma_i(\cdots)$ is an arbitrary state on
$\calA_{{\Omega_i}^c}$.
By using linearlity, $\tilde{\rho}_i(\cdots)$ extends over
whole $\calA$.
The state $\sigma_i(\cdots)$ may be chosen, for example,
as the trace state defined by
$\sigma_i(\cdots)=\lim_{\Gamma\uparrow{\Omega_i}^c}
{\rm Tr}_{\calH_\Gamma}[\cdots]/{\rm Tr}_{\calH_\Gamma}[1]$.
This choice corresponds to the so-called free boundary conditions.
} as an element of 
$\calE$, the compactness ensures 
that one can take a subsequence 
$\{i(j)\}_{j=1,2,\ldots}\subset\{1,2,\ldots\}$ such that the 
weak-$*$ limit
\begin{equation}
\rho(\cdots):=\lim_{j\toinf}\rho_{i(j)}(\cdots)
\label{weak*limit}
\end{equation}
exists.
In the physicists' language, (\ref{weak*limit}) should be read
\begin{equation}
\rho(A)=\lim_{j\toinf}\rho_{i(j)}(A),
\end{equation}
for each $A\in\calA$.

As in Section~\ref{Sec2}, we let $h_o$ be the local Hamiltonian at the origin 
$o\in\Zd$, which acts on the finite dimensional Hilbert space 
$\bigotimes_{x\in\calS_o}\calH_x$ with the support set $\calS_o$ 
containing $r$ sites.
We also set $h_x=\tau_x(h_o)$, and, for any finite set 
$\Omega\in\Zd$,
\begin{equation}
H_{\Omega}:=\sum_{x\in\Omega}h_x,
\end{equation}
where $\tau_x$ denotes the translation by the lattice vector $x$.

We now describe three different definitions of the set of {\em 
ground states}.
The first definition is standard in mathematical literature, and is
\begin{equation}
\calG_1:=\{\omega\in\calE \vbar 
\omega(A^*[H_{\barOmega},A])\ge0
\quad\mbox{for any $A\in\calA_\Omega$, and for any finite 
$\Omega\subset\Zd$}\}.
\end{equation}
Here we introduced
\begin{equation}
\barOmega:=\{x \vbar \calS_x\cap\Omega\ne\emptyset\},
\end{equation}
where $\calS_x=\tau_x(\calS_o)$ is the support set for $h_x$.
(We use the same symbol $\tau_x$ to denote
 the translation operators for 
subsets of $\Zd$ and that for operators.)

The second definition is due to Aizenman and Lieb 
\cite{AizenmanLieb}.
(See also \cite{AizenmanDaviesLieb}.)
The definition is useful because of its similarity to ``classical'' 
definitions of ground states.
It is
\begin{equation}
\calG_2:=\{\omega\in\calE \vbar 
\omega(H_{\barOmega})\le\omega(T(H_{\barOmega}))
\quad\mbox{for any $T\in\calP_\Omega$, and for any finite 
$\Omega\subset\Zd$}\},
\end{equation}
where $\calP_\Omega$ is the set of all local perturbations on 
$\Omega$.
A local perturbation $T$ on $\Omega$ is a linear mapping 
$T:\calA\to\calA$ which satisfies $T(A)\ge0$ for any $A\ge0$, and 
$T(A)=A$ for any $A\in\calA_{\Omega^{\rm c}}$, where 
$\calA_{\Omega^{\rm 
c}}:=\overline{\bigcup_{\Omega'}\calA_{\Omega'\backslash\Omega}}
$ 
is the operator algebra outside of $\Omega$.

The third definition already appeared in Sections~\ref{SecEx} and \ref{SecIG},
and is probably the simplest among the three definitions.
(Essentially the same definition can be found in 
\cite{AffleckKennedy}.)
It is
\begin{equation}
\calG_3:=\{\omega\in\calE \vbar \omega(h_x)=\epsilon_0
\quad\mbox{for any $x\in\Zd$}\},
\end{equation}
where the ground state energy density $\epsilon_0$ is defined as 
follows.
Let $\Lambda$ be the $d$-dimensional $L\times\cdots\times L$ 
hypercubic lattice.
We define the corresponding Hamiltonian with periodic boundary 
conditions as
\begin{equation}
\hampbc=\sum_{x\in\Lambda}h_x,
\label{Hampbc}
\end{equation}
where, for a site $x$ close to the boundary of $\Lambda$, we 
identify $h_x$ 
in (\ref{Hampbc}) as an operator in $\calA_\Lambda$ by imposing 
periodic boundary conditions.
Note that, in the present paper, 
a Hamiltonian with periodic boundary conditions is 
simply denoted as $\ham$ except in the present Appendix.
Then we define $\epsilon_0$ by
\begin{equation}
\epsilon_0:=\lim_\TDL\inf_{\rho_\Lambda\in\calA_\Lambda}
\frac{1}{|\Lambda|}\rho_\Lambda(\hampbc),
\label{DEFe0}
\end{equation}
where $|\Lambda|$ is the number of sites in $\Lambda$, and
the existence of the limit can be proved by a standard 
argument.

For each $i=1,2,3$, we denote by $\hat{\calG}_i$ the set of 
$\omega\in\calG_i$ which is translation invariant, {\em i.e.\/}, 
$\omega(\tau_x(A))=\omega(A)$ for any $A\in\calA$ and any 
$x\in\Zd$.

Now we discuss the relations between these different definitions.
We first note the following.
\begin{pro}
We have $\calG_1=\calG_2$.
\label{G1=G2}
\end{pro}
\begin{proof}{Outline of proof}
Nontrivial parts of the proof is worked out in literature, and we only 
have to make some formal observations.
We make use of the results summarized as Theorem~6.2.52 in 
\cite{BratteliRobinson}.

To prove $\calG_1\subset\calG_2$, we note that the above 
mentioned theorem in \cite{BratteliRobinson} says that 
$\omega\in\calG_1$ if and only if
\begin{equation}
\omega(H_{\barOmega})\le\omega'(H_{\barOmega}),
\end{equation}
for any $\omega'\in\calE$ such that $\omega(B)=\omega'(B)$ for all 
$B\in\calA_{\Omega^{\rm c}}$, and for any finite $\Omega\in\Zd$.
By choosing the perturbed state $\omega'(\cdots)$ in a special form 
$\omega(T(\cdots))$ as in the definition of $\calG_2$, we see that 
$\omega\in\calG_2$.

To prove $\calG_2\subset\calG_1$, we can follow the part (1) 
$\Rightarrow$ (2) of the proof of the above mentioned theorem in 
\cite{BratteliRobinson} without any essential modifications.%
\end{proof}

Next we note that
\begin{pro}
We have $\calG_1=\calG_2\supset\calG_3$.
\label{G3}
\end{pro}
\begin{proof}{Proof}
Because of Proposition~\ref{G1=G2}, it suffices to show 
$\calG_3\subset\calG_2$.
The proof is elementary.

We want to get a contradiction out of the assumption that there is a 
state $\omega$ such that $\omega\in\calG_3$ and 
$\omega\not\in\calG_2$.
From the assumption there exists a finite set $\Omega\subset\Zd$, 
a local perturbation $T\in\calP_\Omega$, and a constant 
$\varepsilon>0$ such that
\begin{equation}
\omega(H_{\barOmega})-\omega(T(H_{\barOmega}))\ge\varepsilon.
\label{omegaT}
\end{equation}
Let $\ell$ be an integer such that $\barOmega$ is contained in a 
suitable $d$-dimensional $\ell\times\cdots\times\ell$ hypercubic 
lattice $\Lambda_0$.
For an integer $n$, let $\Lambda$ be the $d$-dimensional 
$(n\ell)\times\cdots\times(n\ell)$ hypercubic lattice.
There are translation operators $\tau_i$ with $i=1,2,\ldots,n^d$ 
such that
\begin{equation}
\Lambda=\bigcup_{i=1}^{n^d}\tau_i(\Lambda_0).
\end{equation}
Let $\omega_\Lambda$ be the state obtained by simply restricting 
$\omega$ onto $\calA_\Lambda$.
We further define
\begin{equation}
\omega'_\Lambda(A):=\omega_\Lambda(\tilde{T}(A)),
\end{equation}
for a suitable $A\in\calA_\Lambda$, where
\begin{equation}
\tilde{T}=\prod_{i=1}^{n^d}{\tau_i}^{-1}T\tau_i.
\end{equation}
By using (\ref{omegaT}) and the properties of local perturbations, 
we 
observe that
\begin{equation}
\omega_\Lambda(\hampbc)-\omega'_\Lambda(\hampbc)
=\sum_{i=1}^{n^d}\cbk{\omega_\Lambda(H_{\tau_i(\barOmega)})
-\omega_\Lambda(T(H_{\tau_i(\barOmega)}))}
\ge n^d\varepsilon.
\label{bulk}
\end{equation}
On the other hand, since we have $\omega(h_x)=\epsilon_0$, we get
\begin{equation}
\omega_\Lambda(\hampbc)\le
(n\ell)^d\epsilon_0+\beta(n\ell)^{d-1},
\label{surface}
\end{equation}
where $\beta$ is a finite constant which takes care of the boundary 
effects.
By combining the bounds (\ref{bulk}) and (\ref{surface}), we get
\begin{equation}
(n\ell)^{-d}\omega'_\Lambda(\hampbc)-\epsilon_0
\le -\ell^{-d}\varepsilon+\beta(n\ell)^{-1},
\end{equation}
which contradicts with the definition (\ref{DEFe0})
of $\epsilon_0$ by taking $n$ 
sufficiently large.%
\end{proof}

It should be noted that we do not have $\calG_1=\calG_2=\calG_3$.
For example, a state with a single domain wall
in the Ising model belongs to 
$\calG_1=\calG_2$ but not to $\calG_3$.
It is a delicate problem to decide which definition is more 
``realistic''.

As for the translation invariant ground states, however, we have the 
following rather satisfactory result.
\begin{pro}
We have $\hat{\calG}_1=\hat{\calG}_2=\hat{\calG}_3$.
\end{pro}
\begin{proof}{Proof}
Because of Propositions~\ref{G1=G2} and \ref{G3}, it suffices to 
show that $\hat{\calG}_1\subset\calG_3$.
Again the most essential part can be found in literature.
In Theorem~6.2.58 of \cite{BratteliRobinson}, it is proved that a 
translation invariant state $\omega$ belongs to
 $\hat{\calG}_1$ if and 
only 
if $\omega(h_x)=\tilde{\epsilon}_0$ for any $x\in\Zd$.
The ground state energy density is defined as 
\begin{equation}
\tilde{\epsilon}_0:=\inf_{\omega'\in\calE_{\rm inv}}\omega'(h_x),
\end{equation}
where $\calE_{\rm inv}$ is the set of translation invariant states in 
$\calE$.
We only have to show that $\epsilon_0=\tilde{\epsilon}_0$,
 and this may be done in several ways.
Here we offer a simple 
constructive proof.
For each $\Lambda$, we can take a ground state 
$\Pz\in\calH_\Lambda$ of the Hamiltonian $\hampbc$, 
which state is invariant under translations that take into account 
the periodic boundary conditions imposed on $\Lambda$.
Define a state $\omega\in\calE$ by the (weak-$*$) limiting 
procedure (\ref{infGSseq}).
By construction, we see that $\omega\in\hat{\calG}_1$.
The above mentioned theorem of \cite{BratteliRobinson} then 
implies that $\omega(h_x)=\tilde{\epsilon}_0$.
On the other hand, our definition (\ref{DEFe0})
of $\epsilon_0$ implies that 
$\omega(h_x)=\epsilon_0$.%
\end{proof}

One might be interested to know if there is any general theorem 
which tells us exactly what are the elements of the above sets of 
ground 
states.
The following is an example of such general theorems.
It establishes  uniqueness of the ground state when there
is a  decoupled Hamiltonian with a unique
ground state, and then one adds a weak (but completely
arbitrary) translation invariant perturbation to the model.

Suppose that the Hamiltonian at the origin can be written as
\begin{equation}
h_o=v_o+\delta p_o.
\end{equation}
The main part $v_o$ acts only on the space $\calH_o$, and its 
lowest eigenvalue is simple.
The perturbation $p_o$ is an arbitrary self-adjoint operator on 
$\bigotimes_{x\in\calS_o}\calH_x$, and $\delta$ is a constant.
By using a rigorous perturbation technique, the following was proved 
in \cite{KennedyTasaki}.
\begin{theorem}
There exists a finite constant $\delta_0>0$ which depends on the 
dimension $d$, and on the operators $v_o$ and $p_o$.
For $|\delta|\le\delta_0$, the set of ground states 
$\hat{\calG}_1=\hat{\calG}_2=\hat{\calG}_3$ consists of a unique 
element.
\label{UniqueGS}
\end{theorem}

For example the Ising model under sufficiently large transverse 
magnetic field (\ref{Ising1}) is covered by the above theorem by 
setting $h_o=S^{(1)}_o$.
\Section{Ergodic infinite volume ground state in systems with 
discrete symmetry breaking}
\label{APergodic}
In the present Appendix, we concentrate on a system in which 
a discrete symmetry is spontaneously broken.
We assume, for each finite system,
the existence of an ``obscured
symmetry breaking'' and the existence of an energy gap above the
first ``low-lying eigenstate''.
Then we can prove that, by forming a linear combination of the
(finite volume) ground state and the ``low-lying state'', and then
taking an infinite volume limit, one indeed gets an
ergodic infinite volume ground state.

As far as we know this is the first rigorous and general 
result which explicitly
tells one how to construct an ergodic infinite volume ground state when
there is a symmetry breaking.
The theorem is desirable in this sense, but we have to note that
the assumption on the existence of a gap is a rather strong one,
which is not at all easy to verify even 
in relatively simple problems\footnote{
Even in models (like the transverse Ising model of Section~\ref{SecEx})
where one has a convergent cluster expansion, it may not be
easy to verify the existence of the gap.
As for the transverse Ising model in one dimension, one
can make use of the mapping to the free fermion problem to
control the gap.
}.
We also stress that the techniques involved here crucially depend
on the fact that there is only one ``low-lying
eigenstate''.
To prove 
the corresponding conjecture (stated in
Section~\ref{SecIG}) for the models with broken continuous symmetry 
seems formidably difficult at present.

We study the situation basically identical to that in 
Section~\ref{SecHV}, but with additional assumption on the translation
invariance.
The translation invariance is by no means essential in
proving the main theorem, but the implication of the theorem
is interesting only in translation invariant systems.

Let $\Lambda$ be a $d$-dimensional 
hypercubic lattice with periodic boundary conditions, and
denote by $N$ the number of sites in $\Lambda$.
We consider a quantum many-body system on $\Lambda$ as in
Sections~\ref{SecPre} and \ref{SecHV}.
The Hilbert space is constructed as in (\ref{Hilb}), the
Hamiltonian as (\ref{ham}), and the order operator as
(\ref{ord}).
The additional assumptions are that we have
$h_x=\tau_x(h_o)$ and $o_x=\tau_x(o_o)$, where $\tau_x$ is the
translation operator that takes into account the periodic
boundary conditions.
We also require that each $o_x$ acts only on the local Hilbert
space $\calH_x$.
In some situations, one might need to redefine the notion of
``sites'' to satisfy the translation invariance.
See Section~\ref{SecHAF}.

Let $E_\Lambda^{(0)}$, $E_\Lambda^{(1)}$ 
with $E_\Lambda^{(0)}<E_\Lambda^{(1)}$
be the two lowest
eigenvalues of $H_\Lambda$, and $\PLO$, $\PL^{(1)}$ be the 
corresponding normalized eigenstates.
We assume that if $E'_\Lambda$ is any other eigenvalue of
$H_\Lambda$, we have
\begin{equation}
E'_\Lambda-E_\Lambda^{(1)}\ge\EG
\label{gapCond}
\end{equation}
with a ($\Lambda$-independent) constant $\EG>0$.
We also assume that the ground state $\PLO$ exhibits an
``obscured symmetry breaking'' as
\begin{equation}
(\PLO,\op\,\PLO)=0,
\label{Ohat1}
\end{equation}
\begin{equation}
(\PLO,(\op)^2\,\PLO)\ge(\mu oN)^2,
\label{LROap}
\end{equation}
with a constant $\mu>0$, and
\begin{equation}
(\PLO,(\op)^3\,\PLO)=0.
\label{Ohat3}
\end{equation}
Although we did not assume the condition (\ref{Ohat3}) in Section~\ref{SecEx},
it is  valid in most situations.

We shall again consider the ``low-lying state'' of Horsch and von der
Linden \cite{HorschLinden}
\begin{equation}
\Psi_\Lambda:=\frac{\op\PLO}{\norm{\op\PLO}},
\label{Psiap}
\end{equation}
and its linear combination with the ground state
\begin{equation}
\XL:=\frac{1}{\sqrt{2}}\rbk{\PLO+\Psi_\Lambda},
\label{XiAp}
\end{equation}
which was first considered by Kaplan, Horsch, and von der Linden
\cite{KaplanHorschLinden}.
From a straightforward calculation using the 
definitions (\ref{Psiap}), (\ref{XiAp}), and the
assumed (\ref{Ohat1}), (\ref{LROap}), and (\ref{Ohat3}), we find
\cite{KaplanHorschLinden} that
the above state (\ref{XiAp}) exhibits a symmetry breaking as
\begin{eqnarray}
(\XL,\op\,\XL)&=&\frac{1}{2}
\cbk{(\PLO,\op\,\PLO)
+\frac{2(\PLO,(\op)^2\,\PLO)}{\norm{\op\PLO}}
+\frac{(\PLO,(\op)^3\,\PLO)}{\norm{\op\PLO}^2}
}
\ret
&=&\sqrt{(\PLO,(\op)^2\,\PLO)}\ge \mu o N.
\label{OrderAP}
\end{eqnarray}

Let $A$ be a local self-adjoint operator which acts on 
$\bigotimes_{x\in\calS'}\calH_x$, where the 
number of sites in the support set $\calS'$ is bounded by 
a constant $r'$.
For a subset $\Omega\subset\Lambda$, we set
\begin{equation}
\AO:=\sum_{x\in\Omega}\tau_x(A),
\end{equation}
where $\tau_x(A)$ is a translate of $A$ by a lattice vector $x$.
Then the main result of the present Appendix is the following.
\begin{theorem}
We have
\begin{equation}
\lim_{|\Omega|\toinf}\lim_{N\toinf}
\frac{1}{|\Omega|^2}\cbk{(\XL,\AOS\,\XL)-(\XL,\AO\,\XL)^2}
=0,
\label{nofluc}
\end{equation}
for any local operator $A$.
\label{ergodicth}
\end{theorem}

From the Definition~\ref{DEFergodic} of ergodic state, we get the 
following interesting conclusion.
\begin{coro}
The infinite volume ground state
\begin{equation}
 \omega_+(\cdots)=\lim_{N\toinf}(\XL,(\cdots)\,\XL),
\end{equation}
defined by taking a suitable subsequence, is an ergodic translation invariant
ground state.
\label{ergodiccoro}
\end{coro}
It is obvious that the same is true for the infinite volume ground state
$\omega_-(\cdots)$ constructed from $(\PLO-\PsL)/\sqrt{2}$
instead of (\ref{XiAp}).

In the following we prove Theorem~\ref{ergodicth}.
For simplicity, we drop the subscript $\Lambda$ from
$\PLO$, $\PL^{(1)}$, $\PsL$, $\ham$, $\op$, etc.

We start from the following lemma which provides us with the basic
tool in the proof.
In short the lemma says that the set of two states $\{\Phi^{(0)}, 
\Phi^{(1)}\}$ can be used almost as a ``complete basis'' in some situations.
\begin{lemma}
Let $B$ and $C$ be arbitrary self-adjoint operators.
Then for $i,j=0$ or $1$, we have
\begin{eqnarray}
 &&\abs{\amp{i}{BC}{j}
 -\sum_{k=0,1}\amp{i}{B}{k}\amp{k}{C}{j}}
 \ret
 &=&\abs{\amp{i}{B\calP C}{j}}
 \ret
 &\le&\frac{\sqrt{\norm{[B,[H,B]]}\norm{[C,[H,C]]}}}{2\EG},
 \label{EGbound}
\end{eqnarray}
where $\calP$ is the projection operator onto the space orthogonal
to both $\Phi^{(0)}$ and $\Phi^{(1)}$.
\end{lemma}
\begin{proof}{Proof}
From the existence of a gap as in (\ref{gapCond}), we get
the operator inequality
$\calP\le(H-E_i)/\EG$ for $i=0,1$.
By using the Schwartz inequality, we have
\begin{eqnarray}
 &&\abs{\amp{i}{B\calP C}{j}}^2
 \ret
 &\le&\amp{i}{B\calP B}{i}\amp{j}{C\calP C}{j}
 \ret
 &\le&\amp{i}{B\frac{H-E_i}{\EG}B}{i}
 \amp{j}{C\frac{H-E_i}{\EG}C}{j}
 \ret
 &=&(2\EG)^{-2}\amp{i}{[B,[H,B]]}{i}\amp{j}{[C,[H,C]]}{j}
 \ret
 &\le&(2\EG)^{-2}\norm{[B,[H,B]]}\norm{[C,[H,C]]},
\end{eqnarray}
which is the desired bound.%
\end{proof}

As the first application of the lemma, we state the following result
which is both useful and important.
The lemma says that the ``low-lying state'' (\ref{Psiap}) is indeed a very good
approximation of the first excited state $\Phi^{(1)}$.
\begin{lemma}
One can redefine the (quantum mechanical)
phase of the first excited state $\Phi^{(1)}$
so that the bound
\begin{equation}
 \norm{\Phi^{(1)}-\Psi}^2\le\frac{4hr^2}{\EG\mu^2}\frac{1}{N}
 \label{1=PsiA}
\end{equation}
holds.
\label{1=PsiLemma}
\end{lemma}
\begin{proof}{Proof}
Since $(\Phi^{(0)},\Psi)=0$, we can write $\Psi=\alpha\Phi^{(1)}+\Psi'$,
where $\Psi'=\calP\Psi$.
By redefining the phase of $\Phi^{(1)}$, we can choose $\alpha\ge0$.
First note that
\begin{eqnarray}
 \norm{\Phi^{(1)}-\Psi}^2&=&
 \norm{(1-\alpha)\Phi^{(1)}-\Psi'}^2
 \ret
 &=&(1-\alpha)^2+\norm{\Psi'}^2\le2\norm{\Psi'}^2,
\end{eqnarray}
where we have used
$(1-\alpha)^2\le1-\alpha\le1-\alpha^2=\norm{\Psi'}^2$.
To bound $\norm{\Psi'}$, we use (\ref{EGbound}) to get
\begin{eqnarray}
 \norm{\Psi'}^2&=&(\Psi,\calP\,\Psi)
 =\frac{\amp{0}{O\calP O}{0}}{\amp{0}{O^2}{0}}
 \ret
 &\le&\frac{\norm{[O[H,O]]}}{2\EG\amp{0}{O^2}{0}}
 \ret
 &\le&\frac{4ho^2r^2N}{2\EG(o\mu N)^2}
 =\frac{2hr^2}{\EG\mu^2}\frac{1}{N},
\end{eqnarray}
where we used (\ref{LROap}).%
\end{proof}


We now turn to the estimate of the left-hand side of (\ref{nofluc}).
Note that we can assume
\begin{equation}
 \amp{0}{\AO}{0}=0,
 \label{0A0=0}
\end{equation}
since otherwise we can redefine $A-\amp{0}{A}{0}$ as a new $A$.
Let
\begin{equation}
 \Xi':=\frac{1}{\sqrt{2}}(\Phi^{(0)}+\Phi^{(1)}),
 \label{Xi'}
\end{equation}
which is essentially the same as $\Xi$ according to the definition
(\ref{XiAp}) and the relation (\ref{1=PsiA}).
In particular, we have
\begin{equation}
 \absbar\cbk{(\Xi,\AOS\,\Xi)-(\Xi,\AO\,\Xi)^2}
 -\cbk{(\Xi',\AOS\,\Xi')-(\Xi',\AO\,\Xi')^2}\absbar
 \le\frac{a_1\norm{A}^2}{\sqrt{N}}|\Omega|^2.
 \label{XivsXi'}
\end{equation}
Throughout the present proof, $a_i$ denote constants which depend
only on $h$, $r$, $\mu$, and $\EG$.
By using the definition (\ref{Xi'}) and the requirement
(\ref{0A0=0}), we observe that 
\begin{eqnarray}
 &&(\Xi',\AOS\,\Xi')-(\Xi',\AO\,\Xi')^2
 \ret
 &=&\frac{1}{2}\cbk{\amp{0}{\AOS}{0}+\amp{0}{\AOS}{1}
 +\amp{1}{\AOS}{0}+\amp{1}{\AOS}{1}}
 \ret
 &-&\frac{1}{4}\cbk{\amp{0}{\AO}{1}+\amp{1}{\AO}{0}
 +\amp{1}{\AO}{1}}^2
 \ret
 &=&\frac{1}{2}\cbk{\amp{0}{\AOS}{0}
 -\amp{0}{\AO}{1}\amp{1}{\AO}{0}}
 \ret
 &+&\frac{1}{2}\cbk{\amp{0}{\AOS}{1}
 -\amp{0}{\AO}{1}\amp{1}{\AO}{1}}
 \ret
 &+&\frac{1}{2}\cbk{\amp{1}{\AOS}{0}
 -\amp{1}{\AO}{1}\amp{1}{\AO}{0}}
 \ret
 &+&R.
 \label{=R}
\end{eqnarray}
The remaining term $R$ can be further rewritten as
\begin{eqnarray}
 R&=&\frac{1}{2}\amp{1}{\AOS}{1}
 -\frac{1}{4}\cbk{\amp{0}{\AO}{1}^2+\amp{1}{\AO}{0}^2
 +\amp{1}{\AO}{1}^2}
 \ret
 &=&\frac{1}{2}\cbk{\amp{1}{\AOS}{1}
 -\sum_{i=0,1}\amp{1}{\AO}{i}\amp{i}{\AO}{1}}
 \ret
 &+&\frac{1}{4}\cbk{\amp{1}{\AO}{0}\amp{0}{\AO}{1}
 -\amp{1}{\AO}{0}^2}
 \ret
 &+&\frac{1}{4}\cbk{\amp{1}{\AO}{0}\amp{0}{\AO}{1}
 -\amp{0}{\AO}{1}^2}
 \ret
 &+&\frac{1}{4}\amp{1}{\AO}{1}^2.
 \label{R=}
\end{eqnarray}
By using the ``completeness'' relation
(\ref{EGbound}) and (\ref{0A0=0}) to bound the right-hand sides of 
(\ref{=R}) and (\ref{R=}), we have
\begin{eqnarray}
 &&\abs{(\Xi',\AOS\,\Xi')-(\Xi',\AO\,\Xi')^2}
 \ret
 &\le&\frac{1}{2}\abs{\amp{1}{\AO}{0}\amp{0}{\AO}{1}
 -\amp{1}{\AO}{0}^2}
 \ret
 &+&\frac{1}{4}\amp{1}{\AO}{1}^2
 \ret
 &+&\frac{1}{\EG}\norm{[\AO,[H,\AO]]}.
 \label{fluc1}
\end{eqnarray}

We shall bound each term in the right-hand side of (\ref{fluc1}).
To bound the first term, we use (\ref{1=PsiA}) to get
\begin{eqnarray}
 &&\abs{\amp{1}{\AO}{0}\amp{0}{\AO}{1}-\amp{1}{\AO}{0}^2}
 \ret
 &\le&\norm{\AO}\abs{\amp{0}{\AO}{1}-\amp{1}{\AO}{0}}
 \ret
 &\le&\norm{\AO}\cbk{\abs{
 \frac{\amp{0}{(AO-OA)}{0}}{\amp{0}{O^2}{0}^{1/2}}}
 +\frac{a_2\norm{\AO}}{\sqrt{N}}}
 \ret
 &\le&\frac{\norm{\AO}\norm{[\AO,O]}}{\amp{0}{O}{0}^{1/2}}
 +\frac{a_2\norm{\AO}^2}{\sqrt{N}}
 \ret
 &\le&\frac{2\norm{A}^2r'|\Omega|^2}{\mu N}
 +\frac{a_2\norm{A}^2|\Omega|^2}{\sqrt{N}},
 \label{first}
\end{eqnarray}
where we have used the lower bound (\ref{LROap}) and the bound
$\norm{[\AO,O]}\le2\norm{A}or'|\Omega|$.

To bound the second term, we first use (\ref{1=PsiA}) and 
(\ref{LROap}) to get
\begin{eqnarray}
 \amp{1}{\AO}{1}&\le&\frac{\amp{0}{O\AO O}{0}}{\amp{0}{O^2}{0}}
 +\frac{a_3\norm{\AO}}{\sqrt{N}}
 \ret
 &\le&\frac{\amp{0}{O\AO O}{0}}{(\mu o N)^2}
 +\frac{a_3\norm{A}|\Omega|}{\sqrt{N}}.
 \label{1A1}
\end{eqnarray}
To bound the right-hand side of (\ref{1A1}), we use the ``completeness''
relation (\ref{EGbound}) and (\ref{0A0=0}) to get
\begin{eqnarray}
 &&\abs{\amp{0}{O\AO O}{0}}
 \le\abs{\amp{0}{O^2\AO}{0}}+\abs{\amp{0}{O[\AO,O]}{0}}
 \ret
 &&\le\abs{\amp{0}{O^2}{1}\amp{1}{\AO}{0}}
 +\frac{\sqrt{\norm{[O^2,[H,O^2]]}\norm{[\AO,[H,\AO]]}}}{2\EG}
 +\norm{O[\AO,O]}
 \ret
 &&\le\abs{\amp{0}{O^2}{1}\norm{\AO}}
 +\frac{\sqrt{16o^4h^2rN^3\times4\norm{A}^2hrr'^2|\Omega|}}{2\EG}
 +\norm{A}o^2r'|\Omega|N.
 \label{OAO}
\end{eqnarray}
We further use (\ref{1=PsiA}) to see
\begin{eqnarray}
 \abs{\amp{0}{O^2}{1}}&\le&
 \frac{\abs{\amp{0}{O^2O}{0}}}{\amp{0}{O^2}{0}^{1/2}}
 +\frac{a_2\norm{O^2}}{\sqrt{N}}
 \ret
 &\le&o^2a_2N^{3/2},
 \label{0OO1}
\end{eqnarray}
where we used (\ref{Ohat3}).
By substituting (\ref{OAO}) and (\ref{0OO1}) into (\ref{1A1}), we get
\begin{equation}
 \amp{1}{\AO}{1}^2\le c\frac{|\Omega|^2}{N},
 \label{1A1final}
\end{equation}
where $c$ is an $N$-independent constant.

In order to control the third term in the right-hand side of (\ref{fluc1}),
we note that
$\norm{[\AO,[H,\AO]]}\le4\norm{A}^2hrr'^2|\Omega|$.
By putting (\ref{XivsXi'}), (\ref{fluc1}), (\ref{first}),
and (\ref{1A1final}) together, we finally see that
\begin{equation}
 \frac{1}{|\Omega|^2}\abs{(\Xi,\AOS\,\Xi)-(\Xi,\AO\,\Xi)^2}
 \le \frac{4\norm{A}^2hrr'^2}{\EG}\frac{1}{|\Omega|}+O(N^{-1/2}).
\end{equation}
\Section{Lower bound for fluctuation of bulk quantities}
\label{APfluc}
In the present Appendix, we prove simple lemmas which characterize
the behavior of 
fluctuation of bulk quantities in a translation invariant state.
The lemma was used in Sections~\ref{SecEx} and \ref{SecIG} to demonstrate that
the infinite volume ground state obtained as a
limit of finite volume ground states is not ergodic.

Let $\Lambda$ be a $d$-dimensional hypercubic
lattice with $N$ sites and with periodic boundary conditions. 
We consider a quantum many-body system on $\Lambda$ with
the Hilbert space (\ref{Hilb}).
We do not make any specific assumption about the system.
We denote by $\tau_x$ the translation which acts on the operators
and which respects the periodic boundary conditions.

Let $B$ be an arbitrary local operator.
For a subset $\Omega\subset\Lambda$, we set
\begin{equation}
 B_\Omega:=\sum_{x\in\Omega}\tau_x(B).
\end{equation}
\begin{lemma}
Let $\PL$ be an arbitrary state which defines a translation invariant
expectation values, i.e., $\bkt{\tau_x(A)}=\bkt{A}$ for any
local operator $A$.
Then for any local operator $B$, we have
\begin{equation}
 \frac{1}{|\Omega|^2}\bkt{\BO^*\BO}
 \ge\frac{1}{N^2}\bkt{\BL^*\BL}.
 \label{flucBound}
\end{equation}
\label{flucLemma1}
\end{lemma}

Although we only apply the inequality to ground states in the 
present paper, we note that it has a trivial extension to
finite temperature Gibbs states as follows.
We note that the following result has been (implicitly) quoted
in the introduction of 
our previous publication \cite{KomaTasaki}, when we mentioned
that the naive infinite volume limit of
the Gibbs states without
symmetry breaking is not ergodic.
\begin{lemma}
Let $\ham$ be a translation invariant Hamiltonian.
Then for any local operator $B$, we have
\begin{equation}
 \frac{1}{|\Omega|^2}Z(\beta)^{-1}{\rm Tr}
 \sbk{\BO^*\BO e^{-\beta\ham}}
 \ge
 \frac{1}{N^2}Z(\beta)^{-1}{\rm Tr}
 \sbk{\BL^*\BL e^{-\beta\ham}},
\end{equation}
where the partition function is
$Z(\beta)={\rm Tr}\sbk{\exp[-\beta\ham]}$.
\label{flucLemma2}
\end{lemma}

We now prove  Lemma~\ref{flucLemma1}.
Let $\calP_B$ be the projection operator onto the state
$\BL\PL/\norm{\BL\PL}$.
If the state is vanishing, we set $\calP_B=0$.
Since $1-\calP_B$ is nonnegative, we see that
\begin{eqnarray}
 \bkt{\BO^*\BO}
 &\ge&\bkt{\BO^*\calP_B\BO}
 =\frac{\bkt{\BO^*\BL}\bkt{\BL^*\BO}}{\bkt{\BL^*\BL}}
 \ret
 &=&\frac{|\Omega|^2}{N^2}\bkt{\BL^*\BL},
\end{eqnarray}
where we used the translation invariance and the periodic
boundary conditions to get the final line.
If $\calP_B=0$, the inequality is trivial since the
final expression is vanishing.
This proves the lemma.

In order to prove Lemma~\ref{flucLemma2}, we let $\{\Phi^{(n)}\}$
be a complete basis where each basis state $\Phi^{(n)}$ 
is an eigenstate of $\ham$ with the eigenvalue $E_n$, and
also defines translation invariant expectation values.
By using the bound (\ref{flucBound}), we get
\begin{eqnarray}
 Z(\beta)^{-1}{\rm Tr}\cbk{\BO^*\BO e^{-\beta\ham}}
 &=& Z(\beta)^{-1}\sum_n\amp{n}{\BO^*\BO}{n}e^{-\beta E_n}
 \ret
  &\ge& \frac{|\Omega|^2}{N^2}Z(\beta)^{-1}\sum_n
 \amp{n}{\BL^*\BL}{n}
 e^{-\beta E_n}
 \ret
 &=& \frac{|\Omega|^2}{N^2}Z(\beta)^{-1}
 {\rm Tr}\sbk{\BL^*\BL e^{-\beta\ham}}.
\end{eqnarray}
\par\bigskip\bigskip\bigskip
\noindent{\bf Acknowledgments}
\par\noindent
The present work had been under preparation for nearly two years.
We wish to thank 
Ian Affleck,
Tom Kennedy,
Kenn Kubo,
Elliott Lieb,
Seiji Miyashita,
Tsutomu Momoi,
Bruno Nachtergaele, 
Hidetoshi Nishimori,
and
Ken'ichi Takano 
for useful discussions on various related topics.
We also thank Tom Kennedy and Bruno Nachtergaele for valuable comments
on the manuscript.


\end{document}